\documentclass[12pt]{article}
\usepackage{amsmath,amssymb,bm,epsfig}

\renewcommand{\thefootnote}{\fnsymbol{footnote}}

\begin{document}

\title{
\begin{flushright}
\begin{minipage}{0.2\linewidth}
\normalsize
WU-HEP-15-06 \\*[50pt]
\end{minipage}
\end{flushright}
{\Large \bf 
Moduli rolling to a natural MSSM\\ 
with gravitino dark matter
\\*[20pt]}}

\author{Hajime~Otsuka\footnote{
E-mail address: hajime.13.gologo@akane.waseda.jp
}\\*[20pt]
{\it \normalsize 
Department of Physics, Waseda University, 
Tokyo 169-8555, Japan} \\*[50pt]}

\date{
\centerline{\small \bf Abstract}
\begin{minipage}{0.9\linewidth}
\medskip 
\medskip 
\small
We propose the gravitino dark matter in the gravity mediated supersymmetry breaking scenario. 
The mass hierarchies between the gravitino and other superparticles can be achieved by the nontrivial K\"ahler metric of the SUSY breaking field. 
As a concrete model, we consider the five-dimensional supergravity model in which moduli are stabilized, and then one of the moduli induces the slow-roll inflation. 
It is found that the relic abundance of gravitino and the Higgs boson mass reside in the allowed range without a severe fine-tuning. 
\end{minipage}
}

\begin{titlepage}
\maketitle
\thispagestyle{empty}
\clearpage
\tableofcontents
\thispagestyle{empty}
\end{titlepage}

\renewcommand{\thefootnote}{\arabic{footnote}}
\setcounter{footnote}{0}

\section{Introduction}
\label{sec:introduction}
The low-scale supersymmetry (SUSY) is an attractive scenario which not only protects the mass of the Higgs boson from the large radiative corrections but also gives the dark matter candidates. 
In addition to it, the existence of supersymmetry is also motivated in the string theory which is expected as the ultraviolet completions of the standard model (SM). 
This is because the SUSY guarantees the absence of tachyons in the string theory. 

In the minimal supersymmetric standard model (MSSM), the large radiative corrections are indicated by the observed Higgs boson mass within ranges between $124.4$ and $126.8$\,GeV~\cite{Aad:2012tfa}. 
One of the solutions to raise the Higgs boson mass in the MSSM is the high-scale SUSY-breaking scenario, and then the SUSY flavor and {\it CP} problems can be also solved at the same time. 
However, this scenario brings the tuning problem to the MSSM in order to realize the successful electroweak (EW) symmetry breaking. 
By contrast, there is another solution to raise the Higgs boson mass by the nature of left-right mixing of the top squarks. 
As pointed out in Ref.~\cite{Abe:2007kf}, the nonuniversal gaugino masses at the grand unification theory (GUT) scale $M_{\rm GUT}\simeq 2\times 10^{16}$ GeV lead to the maximal mixing of the top squarks and then, 
the realistic Higgs boson mass can be achieved without a severe fine-tuning by the structure of the renormalization group equations in the MSSM. 
Throughout this paper, we focus on this low-scale SUSY breaking scenario. 

The SUSY-breaking scenarios are mainly categorized into the gravity mediation~\cite{Hall:1983iz}, gauge mediation~\cite{Giudice:1998bp} and the anomaly 
mediation~\cite{Randall:1998uk}. 
For any mediation mechanisms, the gravitino mass is sensitive to the cosmological problem, e.g., the cosmological gravitino problem~\cite{Pagels:1981ke}. 
If the gravitino is not stable, the mass of the gravitino should be larger than ${\cal O}(10\,{\rm TeV})$ in order to be consistent with the successful big bang nucleosynthesis (BBN). 
The lower limit of the gravitino mass depends on the reheating temperature; for more details see Refs.~\cite{Pagels:1981ke,Khlopov:1984pf,Moroi:1993mb,Kawasaki:2004qu,Kanzaki:2006hm}.\footnote{It is also possible to consider the light gravitino in the extension of the MSSM. 
See, e.g., Ref.~\cite{Kelso:2013paa}.} Therefore, before discussing our considered situation, we comment on several SUSY-breaking scenarios, focusing on the mass of the gravitino. 

In the gauge mediated SUSY-breaking scenario, the dark matter candidate is the ultralight gravitino of mass $m_{3/2} \ll{\cal O}(1\,{\rm GeV})$ under the low-scale SUSY-breaking. 
Note that if the gravitino mass is larger than this scale, it is expected that the gravitational interactions give the sizable effects to the dynamics of the SUSY-breaking sector as well as the visible sector. 
In the pure anomaly mediated SUSY-breaking scenario, the wino-like neutralino is likely to be the dark matter candidate due to the structure of the beta functions in the MSSM~\cite{Moroi:1999zb}. 
However, the recent results of the LHC experiments~\cite{Aad:2014lra} indicate the TeV scale gluino mass; in other words, the large mass of the gravitino $m_{3/2} \simeq {\cal O}(100\,{\rm TeV})$ is required in the framework of anomaly mediation. 
In the mirage mediation~\cite{Choi:2005uz}, the mixed neutralino would be the dark matter candidate and the large gravitino mass above ${\cal O}(10\,{\rm TeV})$ is expected in the light of the cosmological gravitino problem. 
In the gravity mediation, the neutralino dark matter is often considered under the large gravitino mass above ${\cal O}(10\,{\rm TeV})$ with high-scale SUSY breaking, otherwise, SUSY flavor violations arise due to the flavor dependent interactions. 

In this paper, we consider the gravity mediated SUSY-breaking scenario which is compatible with the low-scale SUSY and observed Higgs boson mass without the cosmological gravitino and SUSY flavor problems. 
In general, the gravity mediation connects the scale of the gravitino mass with that of the supersymmetric particle, because the origin of soft SUSY-breaking terms is only the gravitational interactions.  
Therefore, it seems to be difficult to solve the cosmological gravitino problem with the low-scale SUSY-breaking scenario. 
In order to realize the low-scale SUSY without the cosmological gravitino problem, we propose the mechanism to generate the mass hierarchies between the gravitino and the other sparticles based on the framework of a four-dimensional ${\cal N}=1$ supergravity ($4$D ${\cal N}=1$ SUGRA). 
Especially, we focus on the case that the gravitino is the lightest supersymmetric particle (LSP) whose mass is of O($100$ GeV). 
Since such a stable gravitino is much heavier than that predicted by the gauge mediated SUSY-breaking scenario, this would be the typical feature of the gravity mediation. 
There are some studies for the gravitino dark matter with assumed sparticle spectra that focus on the cosmological implications and it is then found that the next-to-the-lightest supersymmetric particle (NLSP) is severely constrained. 
(See, e.g., Refs.~\cite{Kanzaki:2006hm,Feng:2003xh,Ellis:2003dn,Feng:2004zu}.) 
In order to determine the relevant higher-dimensional operators in $4$D ${\cal N}=1$ SUGRA, we consider a five-dimensional supergravity (5D SUGRA) compactified on an orbifold $S^1/Z_2$.  
In the framework of $5$D SUGRA, the successful inflation mechanism as well as the moduli stabilization can be realized as suggested in Refs.~\cite{Abe:2014vca,Maru:2003mq}. 
The dynamics of inflaton and moduli are important to evaluate the abundance of the gravitino produced via the inflaton and moduli decay into the gravitino. 
Furthermore, the Yukawa hierarchies of elementary particles can be realized without a severe fine-tuning by employing the localized wavefunction of quarks, leptons and Higgs in the fifth dimension~\cite{ArkaniHamed:1999dc}. 

The following sections are organized as follows. 
In Sec.~\ref{sec:N=1SUGRA}, we discuss how to realize the mass hierarchies between the gravitino and other supersymmetric particles in $4$D ${\cal N}=1$ SUGRA. 
As a concrete model, in Sec.~\ref{sec:5DSUGRA}, we briefly review the structure of $5$D SUGRA on $S^1/Z_2$ and then the gravitino can be the dark matter candidate. 
Thanks to the detailed moduli stabilization as well as the inflation mechanism, one can discuss the nonthermal productions of the gravitino via the moduli and inflaton decay after the inflation. 
After that, in Sec.~\ref{sec:gravitino}, we evaluate the relic abundance of the gravitino and the Higgs boson mass with a severe fine-tuning. 
The obtained results are consistent with the cosmological observations as well as the collider experiments. 
Finally, Appendices~\ref{app:oscvac} and~\ref{app:oscinf} denote the detailed derivations of the scalar potential around the vacuum and during the inflation, respectively.

\section{The mass hierarchies between the gravitino and other sparticles}
\label{sec:N=1SUGRA}
In this section, we show how to realize the mass hierarchies between the gravitino and other sparticles in the framework of $4$D ${\cal N}=1$ SUGRA. 
The scalar potential in $4$D ${\cal N}=1$ SUGRA is given by 
\begin{align}
V &=e^{K}\left( K^{I\bar{J}} D_{I}W D_{\bar{J}}\bar{W} -3|W|^2\right)
\nonumber\\
   &=K_{I\bar{J}} F^{I} F^{\bar{J}} - 3e^{K/2}|W|^2,
\label{eq:scalarpo}
\end{align}
where $K$ and $W$ are the K\"ahler and superpotential, respectively. $D_I W =W_I + K_I W$ with $W_I=\partial W/\partial Q^I$, $K_I=\partial K/\partial Q^I$ 
are the K\"ahler covariant derivatives of the superpotential for the scalar components of the chiral superfields $Q^I$, $F^I=-e^{K/2}K^{I\bar{J}}D_{\bar{J}}\bar{W}$ 
are the {\it F}-terms of $Q^I$ and $K^{I\bar{J}}$ are the inverse of K\"ahler metric $K_{I\bar{J}}=\partial_I\partial_{\bar{J}}K$. 
Here and hereafter, we set the Planck unit $M_{\rm Pl}=1$, unless we specify it. 
The vanishing cosmological constant $\langle V\rangle =0$ is rewritten in the following form:
\begin{align}
m_{3/2}^2=\frac{1}{3}\langle K_{X\bar{X}}F^X F^{\bar{X}}\rangle, 
\label{eq:gravitinomass}
\end{align}
where $m_{3/2} = e^{\langle K\rangle/2} \langle W\rangle$ is the gravitino mass. It is then assumed that the SUSY is broken by the single chiral superfield $X$,\footnote{It is straightforward to extend our situation 
in multiple SUSY-breaking fields.} whereas the soft SUSY-breaking masses of the gauginos and scalar components of the chiral superfields $Q^I$ are given by
\begin{align}
 M_a &= \langle F^X\partial_X \text{ln}\left(\text{Re} f_a \right)\rangle, \nonumber\\
 m_{\Phi^I}^2 &= 
 -\langle F^X\bar{F}^{\bar{X}}\partial_X\partial_{\bar{X}} \text{ln} Y_{\Phi^I}\rangle, 
 \label{eq:softmasses}
\end{align}
where $f_a$, $a=U(1)_Y, SU(2)_L, SU(3)_C$ are the gauge kinetic functions of the standard model gauge 
groups whose vacuum expectation values (VEVs) determine the size of gauge couplings. 
$Y_{Q^I}$ are some nontrivial functions for the kinetic term of $Q^I$ which can be severely constrained by 
the flavor structure of elementary particles as can be seen later. 
From the above Eqs.~(\ref{eq:gravitinomass}) and (\ref{eq:softmasses}), the nontrivial K\"ahler metric 
of the SUSY-breaking field $X$ gives rise to two nontrivial possibilities: 
\begin{description}
\item[{$\bullet$}]  The gravitino dark matter:\\
In the case of $\langle K_{X\bar{X}}\rangle \ll 1$, the gravitino mass is smaller than the soft SUSY-breaking masses for any value of the F-term $\langle F^X\rangle$. 
Then it is possible to consider the gravitino dark matter in the gravity mediated SUSY-breaking scenario with TeV scale gauginos and sparticles. 
It is then assumed that the derivatives of the gauge kinetic function $\partial_X {\rm Re} f_a$ and 
the kinetic term of $\Phi^I$, $\partial_X\partial_{\bar{X}} \text{ln} Y_{\Phi^I}$ satisfy the certain conditions in order to obtain the 
gravitino dark matter. 
This is because the renormalization group (RG) effects are significant to discuss the sparticle spectrum. 
Such conditions are discussed in the case of constrained MSSM (CMSSM)~\cite{Kersten:2009qk}. 

The stable gravitino would be consistent with the thermal history of the universe, even if the decays of NLSP do not spoil the success of 
BBN~\cite{Kanzaki:2006hm,Feng:2003xh,Ellis:2003dn,Feng:2004zu} and at the same time, the relic abundance of the gravitino should not be larger than that reported by the Planck Collaboration~\cite{Ade:2015lrj}. 
In any case, the stable gravitino is favored in the light of naturalness, because the {\it F}-term of the SUSY-breaking field can be taken as a usual low-scale SUSY-breaking scenario which soften the divergences for the Higgs boson mass. 
Note that the small K\"ahler metric of the field $X$ should be ensured in order not to be below that generated by the loop and/or higher derivative corrections. 

\item[{$\bullet$}]  Other dark matter candidates:\\
By contrast, in the case of $\langle K_{X\bar{X}}\rangle \gg 1$, it is expected that the gravitino is heavier than the other sparticles for any value of 
the {\it F}-term $\langle F^X\rangle$ with $\partial_X {\rm Re} f_a$ and 
$\partial_X\partial_{\bar{X}} \text{ln} Y_{\Phi^I}$ of order unity. 
Thus, one can solve the cosmological gravitino problem with the low-scale SUSY-breaking scenario. 
Then, the gravitino mass can be chosen as above $10$ TeV; otherwise, the BBN is threatened by the gravitino decay into the electronic and hadronic showers. 
Although we do not pursue this possibility, it is interesting to work in this direction.\footnote{We will discuss it in the separate work.}  
\end{description}

\section{Gravitino dark matter in 5D SUGRA}
\label{sec:5DSUGRA}
\subsection{4D effective Lagrangian and matter contents}
\label{subsec:4Deff}
The soft SUSY-breaking terms are sensitive to the ultraviolet completion of the SM. 
As a concrete model, we consider the $5$D SUGRA on $S^1/Z_2$ and the flat $5$D background metric,
\begin{equation*}
ds^2 = \eta_{\mu\nu} dx^\mu dx^\nu -dy^2,
\end{equation*}
where $x^\mu$ with $\mu =0, 1, 2, 3$ and $y$ denote the $4$D spacetime and fifth coordinates, respectively. 
$\eta_{\mu\nu} =$diag$(1, -1, -1, -1)$ and the fundamental region of the orbifold is chosen as $0\leq y\leq L$ in which $y=0,L$ correspond to the fixed points. 
The $S^1/Z_2$ orbifold restricts all fields $f(x,y)$ to two classes of them such as $Z_2$-even and -odd fields, satisfying the following $Z_2$ transformations $f(x,-y)=f(x,y)$ and $f(x,-y)=-f(x,y)$, respectively. 
Only $Z_2$-even fields have zero modes which can appear in the low-energy effective theory. 

First of all, we list the relevant matter contents of $5$D SUGRA. 
From the structure of the orbifold, $5$D SUSY is broken into the $4$D ${\cal N}=1$ SUSY. Correspondingly, $5$D vector multiplets ${\bm V}^I$ and hypermultiplets ${\bm \Phi}_\alpha$ are decomposed into $4$D vector multiplets $V^I$ and three types of chiral multiplets $\Sigma^I$, $\Phi_\alpha$ and $\Phi_\alpha^C$, that is, ${\bm V}^I=\{V^I, \Sigma^I\}$ with $I=1,2,\cdots,n_v$ and ${\bm H}_\alpha=\{\Phi_\alpha,\Phi_\alpha^C\}$ with $\alpha=1,2,\cdots,n_{H}+n_C$ where $n_C$ is the number of compensator hypermultiplets and in this paper, it is chosen as $n_C=1$, for simplicity. 
In addition to the usual $Z_2$-even vector multiplets $V^{I}$ involving the vector multiplets in the standard model, we consider $U(1)_{I'}$ $Z_2$-odd vector multiplets ${\bm V}^{I'}$ with $I'=1,2,\cdots, n_V^{I'}$. 
The zero modes of $Z_2$-even chiral multiplets $\Sigma^{I'}$ are called as the moduli chiral multiplets $T^{I'}$ whose linear 
combination\footnote{In the case $n_V^{I'}=1$, the radion multiplet corresponds to the single modulus $T^{I'=1}$.} plays a role of the inflaton field as pointed out in Ref.~\cite{Abe:2014vca}. 
In what follows, we define the zero mode of chiral multiplets $\Phi_\alpha$ as $Q_\alpha$ which involve the quark chiral multiplets $({\cal Q}_i,{\cal U}_i,{\cal D}_i)$, lepton chiral 
multiplets $({\cal L}_i,{\cal E}_i, N_i)$ with the number of generations $i=1,2,3$, Higgs chiral multiplets $({\cal H}_u,{\cal H}_d)$, SUSY-breaking chiral multiplet $X$ and the stabilizer multiplets $H_{I'}$. 
These multiplets have representations of the standard model gauge groups and extra $U(1)_{I'}$ gauge groups whose gauge fields $A_M^{I}$, $A_M^{I'}$ in vector multiplets ${\bm V}^I$ and ${\bm V}^{I'}$, respectively. 
It is then assigned $U(1)_{I'}$ charges $c_{I'}^{(\alpha)}$ to these hypermultiplets ${\bm H}_\alpha$. 
Here it is assumed that the visible sector consists of the MSSM plus right-handed (s)neutrinos and the same number of stabilizer hypermultiplets as that of moduli multiplets in order to generate the moduli and inflaton potential as can be shown later. 

Next, we show the effective action obtained from the $5$D conformal supergravity action for vector and hypermultiplets which is an off-shell description of $5$D SUGRA~\cite{Zucker:1999ej,Kugo:2000af}. 
The structure of the K\"ahler potential in 5D SUGRA on $S^1/Z_2$ can be characterized by the cubic polynomial of vector multiplets, the so-called {\it norm function}, ${\cal N}(M)=\sum_{I,J,K=1}^{n_V} C_{I,J,K}M^IM^JM^K$ with real coefficients $C_{I,J,K}$ for $I,J,K=1,2,\cdots,n_V$, and the $U(1)_{I'}$ charges of hypermultiplets. 
After the off-shell dimensional reduction discussed in Refs.~\cite{Abe:2005ac,Correia:2006pj,Abe:2006eg} based on the $4$D ${\cal N}=1$ 
superspace~\cite{Abe:2004ar,Paccetti:2004ri},\footnote{The more general $5$D action, including $Z_2$-odd fields, is discussed in Refs.~\cite{Sakamura:2011df,Sakamura:2012bj}} the $4$D effective Lagrangian is given by 
\begin{align}
 {\cal L}_{\rm eff} =& -\frac{1}{4}\left[\int d^2\theta ~ 
 \sum_a f_a(X, T)\text{tr}({\cal W}^a{\cal W}^a )+\text{h.c.} \right] +\int d^4\theta ~~ |\phi|^2 \Omega_{\rm eff}(|Q|^2,\text{Re} T)
 \nonumber \\ 
 &+\left[\int d^2\theta ~~ \phi^3 W(Q,T)+\text{h.c.} \right], 
 \label{eq:Leff}
\end{align}
where $\phi$ is the compensator multiplet, ${\cal W}^a$ is the field strength supermultiplet for a massless 4D vector multiplets $V^a$ with $a=U(1)_Y, SU(2)_L, SU(3)_C$ originating 
from the $5$D $Z_2$-even multiplets ${\bm V}^a$, $Q_{\alpha}$ are the $4$D chiral multiplets, $X$ is the $4$D chiral multiplet which induces the SUSY breaking, and $T^{I'}$ are the moduli chiral multiplets. 

Then, the gauge kinetic functions $f_a(X,T)$ in Eq.~(\ref{eq:Leff}) are supposed as 
\begin{equation}
 f_a(X,T) =\xi^a_X X +\sum_{I'=1}^{n_V^{I'}}\xi^a_{I'}T^{I'},
 \label{eq:gaugekin}
\end{equation}
where $\xi_{I'}^a$ and $\xi_{X}^a$ are real constants determined by the real coefficients $C_{I',J,K}$ in the norm function and the gauge kinetic functions at the orbifold fixed point $y=0$, respectively. 
Since the gauge kinetic functions at the orbifold fixed points depend on the dynamics of the SUSY-breaking sector, we comment on the reason why we take the above ansatz later. 

On other hand, the effective K\"ahler potential in Eq.~(\ref{eq:Leff}) is given by
\begin{align}
 \Omega_{\rm eff}(|Q|^2,\text{Re} T) 
 &= {\cal N}^{1/3}(\text{Re} T)\left[
 -3 +2\sum_a Y(c_\alpha \cdot T) |Q_\alpha|^2 +\sum_{\alpha, 
\beta}\tilde{\Omega}^{(4)}_{\alpha ,\beta}(\text{Re} T) 
|Q_\alpha|^2 |Q_\beta|^2 
 +{\cal O}\Bigl( |Q|^6\Bigl) \right], \nonumber\\
 \label{eq:effKahler}
\end{align}
without the K\"ahler potential at the orbifold fixed points $y=0,L$, where ${\cal N} (\text{Re} T)$ is the norm function, $Y(z)\equiv (1-e^{-2\text{Re} z})/2\text{Re} z$ 
stands for the kinetic terms of $Q_{\alpha}$ which have appeared after solving their equation of motion in the fifth direction and $c_\alpha^{I'}$ denote the $U(1)_{I'}$ charges of $Q_\alpha$. The four-point couplings $\tilde{\Omega}^{(4)}_{\alpha, \beta}$ 
are defined as
\begin{eqnarray}
 \tilde{\Omega}^{(4)}_{\alpha ,\beta} &\equiv& -\frac{(c_\alpha
\cdot{\cal P} a^{-1}\cdot c_\beta)
 \{ Y((c_\alpha +c_\beta)\cdot T)-Y(c_\alpha\cdot T)
Y(c_\beta\cdot T) \} }
 {(c_\alpha \cdot \text{Re} T)(c_\beta \cdot \text{Re} T)}
 +\frac{Y((c_\alpha +c_\beta)\cdot T)}{3},\nonumber\\
{\cal P}^{I}_{~~J}({\cal X}) &\equiv& \delta^{I}_{~~J}-\frac{{\cal X}^I{\cal N}_J}{3{\cal N}}({\cal X}),  
\label{eq:Omg4}
\end{eqnarray}
where ${\cal P}^{I}_{~~J}({\cal X})$ is the operator to project the moduli multiplets out the radion multiplet. 
The notable feature there is that the flavor structure of matter fields is characterized by the $U(1)_{I'}$ charges of them in the K\"ahler potential~(\ref{eq:effKahler}). 
By contrast, the superpotential can be allowed only at the orbifold fixed points where the SUSY is reduced to the $4$D ${\cal N}=1$. Therefore, we consider the superpotential including the Yukawa couplings and $\mu$-term in the MSSM, moduli potential at $y=0$, and the moduli potential at $y=L$, respectively. 
The explicit form of the superpotential in Eq.~(\ref{eq:Leff}) is shown later.

\subsection{Gravitino dark matter in $5$D SUGRA}
\label{subsec:gravitino}
In this section, we show the realization of mass hierarchies between the gravitino and other sparticles in the framework of $5$D SUGRA. 
As shown in Eq.~(\ref{eq:effKahler}) in Sec.~\ref{subsec:4Deff}, the bulk K\"ahler potential is rewritten as
  \begin{equation}
K_{\rm bulk}= -{\text ln} {\cal N} (\text{Re} T) + \sum_{a} Z_{Q_a}(\text{Re} T) |Q_a|^2 + Z_X(\text{Re} T) |X|^2 +{\cal O}(|Q|^4), 
\label{eq:bulkK}
  \end{equation}
where the K\"ahler metric $K_{X\bar{X}}$ for the SUSY-breaking field $X$ is given by 
\begin{align}
K_{X\bar{X}} = Z_X(\text{Re}T) &=\frac{1-e^{-2c_X\cdot \text{Re}T}}{c_X\cdot \text{Re}T} \nonumber\\
&\simeq 
\begin{cases}  
\frac{1}{c_X\cdot \text{Re}T},  & c_X\cdot \text{Re}T>0, \\ 
 \frac{1}{|c_X\cdot \text{Re}T}|\text{exp}(2|c_X\cdot \text{Re}T|) & c_X\cdot \text{Re}T<0,
\end{cases}
\label{Kahlermetric} 
\end{align}
where $K_{X\bar{X}}$ depends on the $U(1)_{I'}$ charges of the field $X$ for the $Z_2$-odd vector multiplets ${\bm V}^{I'}$ and the VEVs of the moduli $T^{I'}$, except for the case of the vanishing $U(1)_{I'}$ charges. 
For the mild large volume of the fifth dimension, $L \simeq {\cal N}^{1/2} (\langle\text{Re} T\rangle) \gg 1$ and positive $U(1)_{I'}$ charges, the VEV of the K\"ahler metric 
is smaller than O(1), that is, $\langle K_{X\bar{X}}\rangle \ll 1$, which is important to obtain so that the light gravitino can be lower than the other sparticles. 

The soft SUSY-breaking masses for the scalar components of $Q_\alpha$ are given by the four-point couplings $\tilde{\Omega}^{(4)}_{\alpha ,X}$ in Eq.~(\ref{eq:Omg4}). 
For typical $U(1)_{I'}$ charges of $Q_\alpha$ to realize the realistic Yukawa couplings, the soft SUSY-breaking masses are larger than the gravitino mass as shown later. 
Furthermore, the gauge kinetic functions in Eq.~(\ref{eq:gaugekin}) lead to the following gaugino masses at the compactification scale by employing the formula (\ref{eq:softmasses}), 
\begin{align}
M_a=\frac{F^X}{g_a^2}\xi^a_X +\sum_{I'=1}^{n_V^{I'}} \frac{F^{T^{I'}}}{g_a^2} 
\xi_{I'}^a. 
\end{align}
When the compactification scale is close to the GUT scale, we obtain the gaugino masses at the EW scale after solving the one-loop RG equations from the GUT scale to the EW scale, 
\begin{align}
M_1(M_{\rm EW})\simeq 0.4\,M_1(M_{\rm GUT}), \,\,\,
M_2(M_{\rm EW})\simeq 0.8\,M_2(M_{\rm GUT}),\,\,\,
M_3(M_{\rm EW})\simeq 2.9\,M_3(M_{\rm GUT}). 
\end{align}
Then, the gravitino LSP occurs if these gaugino masses at the EW scale are larger than the gravitino, as pointed out in Ref.~\cite{Kersten:2009qk}. 
In the case of $5$D SUGRA, such situations can be realized by properly 
choosing the parameters $\xi^a_X$, $\xi^a_{I'}$ and at the same time, the Higgsino mass should be larger than the gravitino mass. 
Thus, one can consider the gravitino dark matter in the gravity mediated SUSY-breaking scenario without changing the VEVs of the {\it F}-terms as discussed in Sec.~\ref{sec:N=1SUGRA}. 
In order to estimate thermal and nonthermal abundances of gravitino via the moduli and/or inflaton decay, we focus on the specific model which realizes the successful moduli inflation as well as the moduli stabilization~\cite{Abe:2014vca} in the next section~\ref{subsec:moduli_stabilization}.  

The mild large volume also reduces the contribution from the K\"ahler potential at the orbifold fixed points $y=0, L$ to be small compared with the bulk K\"ahler potential~(\ref{eq:bulkK}). 
Since their boundary terms are described by 
\begin{equation}
K_{\rm boundary}= {\cal N}^{-1/3}\left( K^{(0)}(|X|^2) +K^{(L)}(e^{-c_X\dot (T+\bar{T})}|X|^2) +\cdots \right),
\end{equation}
the overall factor ${\cal N}^{-1/3}$ suppress these contributions. The one-loop corrections to the moduli K\"ahler potential~\cite{Sakamura:2013wia} are also suppressed by the mild large volume of the fifth dimension. 

By contrast, in the case of negative $U(1)_{I'}$ charges, the VEV of the K\"ahler metric is bigger than O(1), that is, $\langle K_{X\bar{X}}\rangle \gg 1$. 
From the mass formula of the gravitino and sparticles given by Eqs.~(\ref{eq:gravitinomass}) and (\ref{eq:softmasses}), one can expect that the sparticles are lighter than 
the gravitino without changing the {\it F}-term of the SUSY-breaking field. 
Thus, it is possible to solve the gravitino and fine-tuning problems at the same time.

\subsection{Moduli stabilization}
\label{subsec:moduli_stabilization}
Following the discussion about the small-field inflation in Ref.~\cite{Abe:2014vca}, we choose the norm function as 
\begin{equation}
{\cal N} (\text{Re} T) =(\text{Re} T^1)(\text{Re} T^2)(\text{Re} T^3),
\end{equation}
which leads to the diagonal moduli K\"ahler metric. 
Because it seems to be difficult to obtain the realistic masses and mixings of quarks and leptons in the case of two moduli as shown in Sec.~\ref{subsec:Yukawa}, we restrict ourselves to the case of three moduli $T^{I'=1,2,3}$ in what follows. 
In order to generate the moduli potential, we introduce the same number of stabilizer chiral multiplets $H_i$ as that of moduli chiral multiplets as stated in Sec.~\ref{subsec:4Deff}. 
The effective K\"ahler potential, except for the SUSY-breaking field $X$ and other matters in the MSSM are 
  \begin{equation}
K= -{\text ln} {\cal N} (\text{Re} T)+ \sum_{i=1}^3 Z_{H_i}(\text{Re}T^{I'=i}) 
|H_i|^2, 
\label{eq:Kmo}
  \end{equation}
where it is then assumed that the stabilizer fields $H_i$ have only the $U(1)_{I'=i}$ charge with $i=1,2,3$, for simplicity. 
In addition to it, the relevant superpotential for the moduli inflation and stabilization is 
\begin{align}
W_{\rm mod} &=\sum_{i=1}^3 J_{0}^{(i)}H_i^{(0)} 
-\sum_{i=1}^3J_{L}^{(i)}H_i^{(L)}\nonumber\\
&=\sum_{i=1}^3 \left( J_{0}^{(i)}-J_{L}^{(i)}e^{-c_{I'}^{(i)}T^{I'}} 
\right) H_i^{(0)},
\label{eq:Wmo}
\end{align}
where $J_{0,L}^{(i)}$ are constants at the orbifold fixed points $y=0,L$ and the exponential factor $e^{-c_{I'}^{(i)}T^{I'}}$ comes from the profile of the wavefunction of the stabilizer fields in the fifth direction, $H_i^{(L)}=e^{-c_{I'}^{(i)}T^{I'}}H_i^{(0)}$. 
Here we assume that these tadpole terms are dominant in the superpotential and the other terms are negligible due to some symmetries or dynamics.\footnote{A similar moduli stabilization was proposed in Ref.~\cite{Abe:2007zv} in the case of $n_C=2$.} 
In the following, we omit the subscripts of the stabilizer fields at the fixed point $y=0$, that is $H_i=H_i^{(0)}$.  

In fact, from the $4$D scalar potential (\ref{eq:scalarpo}) given by the K\"ahler and superpotential (\ref{eq:Kmo}), (\ref{eq:Wmo}), the expectation values of the moduli $T^{I'}$ 
and the stabilizer fields $H_i$ are found as~\cite{Maru:2003mq} 
\begin{align}
c_{I'}^{(i)}\langle T^{I'}\rangle 
=\ln \frac{J_{L}^{(i)}}{J_{0}^{(i)}}, 
\,\,\, \langle H_i\rangle =0,
\label{eq:refmoduli}
\end{align}
which are determined by the stabilization conditions, $\langle D_{I'}W\rangle =\langle D_{i}W\rangle =\langle W\rangle=0$, at which the supersymmetric Minkowski minimum can be realized,~$\langle V\rangle=0$. 
Their supersymmetric masses of moduli and stabilizer fields are estimated as 
\begin{align}
m^2_{I'i} \simeq \frac{e^{\langle K\rangle} 
\langle W_{I'i}\rangle^2 }{\langle K_{I'\bar{I}'} \rangle 
\langle K_{i\bar{i}} \rangle},
\label{eq:SUSYmass}
\end{align}
where $\langle W_{I'i}\rangle =-c_{I'}^{(i)} J_L^{(i)}e^{-c_{I'}^{(i)}T^{I'}}$ and $W_{ij}=\partial_i\partial_j W$. 
Now there are no mixing terms between the moduli and stabilizer fields in the mass matrices due to the diagonal K\"ahler metric of them. 
From the exponential behaviors of supersymmetric masses~(\ref{eq:SUSYmass}), the mass scales of moduli and stabilizer fields are controlled by the sizes of $U(1)_{I'=i}$ charge and constants $J_{0,L}^{(i)}$. 

So far, the SUSY is not broken in the superpotential~(\ref{eq:Wmo}). 
For the SUSY-breaking sector, we consider the O'Raifeartaigh model~\cite{O'Raifeartaigh:1975pr} which is simplified as the following K\"ahler and superpotential of the SUSY-breaking field $X$ after integrating out the heavy modes, 
\begin{align}
K=Z_X(\text{Re}T^1, \text{Re}T^2) |X|^2 
-\cfrac{1}{\Lambda^2} |X|^4,\,\,\,
W=w +\nu X,
\label{eq:KWX}
\end{align}
where $w$, $\nu$ are the real parameters and the SUSY-breaking field $X$ has no $U(1)_3$ charge, for simplicity. The K\"ahler potential receives the loop corrections from the heavy modes, whose mass scale is $\Lambda$~\cite{Kallosh:2006dv}. 

In general, the true vacuum of the moduli and stabilizer fields are deviated from the supersymmetric one due to the SUSY-breaking effects and then the moduli and stabilizer fields obtain 
their {\it F}-terms at the true vacuum. 
Since their {\it F}-terms would change the cosmological history of the universe through the moduli decay into the gravitinos, it is important to evaluate their {\it F}-terms at the true vacuum. For that reason, we adopt the perturbation method, known as the reference point method~\cite{Abe:2006xp} to search for the true vacua of all the fields.

First, as the reference points for the moduli and stabilizer fields, we take them as given in Eq.~(\ref{eq:refmoduli}) satisfying as
\begin{align}
D_{H_i} W|_{\rm ref} = W_{H_i} +K_{H_i} W =0,\,\,\,
D_{T^{I'}} W|_{\rm ref} = K_{T^{I'}} w,
\end{align}
and for the SUSY-breaking field $X$, its reference point is taken as that satisfying the following stabilization condition:
\begin{align}
&e^{-K} V_X|_{\rm ref} = \partial_{X} 
(\sum_{I'} K^{T^{I'} {\bar T}^{\bar{I}'}} |D_{T^{I'}} W|^2+K^{X {\bar X}} |D_{X} W|^2 - 3|W|^2) \nonumber\\
& \hspace{2cm} \simeq  3W_X \bar{W} +\partial_{X} (K^{X {\bar X}}) |W_{X}|^2 + K^{X {\bar X}}W_X K_{X {\bar X}} {\bar W} -3 W_{X} {\bar W} \nonumber\\
& \hspace{2cm} \simeq \cfrac{4|W_{X}|^2}{\Lambda^2 (Z_X)^2}  {\bar X} +W_X {\bar W} = 0,
\end{align}
in the limit $w\ll 1$, where $V_X=\partial_X V$. 
Thus, we obtain 
\begin{align}
X|_{\rm ref} = -\cfrac{\Lambda^2 (Z_X)^2}{4} \left( \cfrac{W}{W_X} \right) \simeq -\cfrac{\Lambda^2 (Z_X)^2 w}{4\nu}.
\label{eq:modulivev}
\end{align}
Next, we expand these fields as $\phi \rightarrow \phi|_{\rm ref}+ \delta \phi$, $\phi=T^{I'}, H_i, X$ with $I',i=1,2,3$ and evaluate their 
perturbations from the reference points given by Eqs.~(\ref{eq:refmoduli}) and (\ref{eq:modulivev}) under the following conditions:
\begin{align}
&V =V|_{\rm ref} +V_I|_{\rm ref} \delta\phi^I +V_{\bar{I}}|_{\rm ref} \bar{\delta\phi^I}+V_{IJ}|_{\rm ref} \delta\phi^I \delta\phi^J +V_{I\bar{J}}|_{\rm ref} \delta\phi^I \bar{\delta\phi^J} +V_{\bar{I}\bar{J}}|_{\rm ref} \bar{\delta\phi^I} \bar{\delta\phi^J} +O(\delta \phi^3), \nonumber \\  
&\Bigl|V_I|_{\rm ref} \delta\phi^I +V_{\bar{I}}|_{\rm ref} \bar{\delta\phi^I} \Bigl| \gg \Bigl|V_{IJ}|_{\rm ref} \delta\phi^I \delta\phi^J +V_{I\bar{J}}|_{\rm ref} \delta\phi^I \bar{\delta\phi^J} +V_{\bar{I}\bar{J}}|_{\rm ref} \bar{\delta\phi^I} \bar{\delta\phi^J}\Bigl|,
\label{refpointcd}
\end{align}
where $V_I=\partial_I V$ and $V_{IJ}=\partial_I\partial_J V$ are the derivatives for the relevant fields $\phi$, and then $\phi|_{\rm ref}+ \delta \phi$ are considered as the vacua of relevant fields. 
Note that these perturbations are valid even if the SUSY-breaking scale is smaller than the scale of supersymmetric masses given by Eq.~(\ref{eq:SUSYmass}). 
As a result, the deviations of the fields from the reference points~(\ref{eq:refmoduli}), (\ref{eq:modulivev}) are 
\begin{align}
&\delta H_i \simeq \frac{w}{2\text{Re}T^{I'} W_{T^{I'}H_i}},\,\,\, \delta T^{I'}\simeq \left(\frac{w}{W_{T^{I'}H_i}}\right)^2,\,\,\,
\delta X\simeq \left(\frac{\Lambda^2 Z_X^2}{4w^2}\right) 5wW_X, 
\label{eq:variation}
\end{align}
and the {\it F}-terms and squared masses of moduli, stabilizer, and SUSY-breaking fields are roughly estimated as
\begin{align}
&\sqrt{K_{T^{I'}{\bar T}^{I'}}}F^{T^{I'}} \simeq O\left(\frac{w^3}{m_{T^{I'}}^2}\right), \;\;  \sqrt{K_{H_i\bar{H}_i}}F^{H_{i}} \simeq O\left(\frac{w^3}{m_{H_i}^2}\right),\;\; \sqrt{K_{X\bar{X}}}F^X \simeq \frac{-\nu}{{\cal N}^{1/2}Z_X^{1/2}} \nonumber\\
&m_{T^{I'}}^2\simeq m_{H_i}^2 \simeq \cfrac{e^K W_{T^{I'}H_i}^2}{K_{T^{I'}{\bar T}^{I'}} K_{H_i{\bar H}_i}}\;\;(I'=i),
\,\,\,\,\, 
m_X^2 \simeq \cfrac{e^K}{K_{X\bar{X}}} \frac{4w^2}{\Lambda^2 Z_X^2},
\label{eq:momasseana} 
\end{align}
at the vacuum, $\phi=\phi|_{\rm ref} + \delta \phi$. The mass squares of 
real and imaginary parts of moduli, stabilizer, and SUSY-breaking fields are the same as each other and here and in what follows, they are denoted as $m_{T^{I'}}^2$, $m_{H_i}^2$ and $m_{X}^2$, respectively. 
The details of these derivations are summarized in Appendix.~\ref{app:oscvac}. 
The mass differences between $m_{T^{I'=i}}$ and $m_{H_i}$ are the order of the gravitino mass. 
It is remarkable that the fields, except for the SUSY-breaking field $X$, have almost vanishing {\it F}-terms due to their large supersymmetric masses.

\subsection{Moduli inflation}
\label{subsec:moduli_inflation}
In this section, we briefly review the inflation mechanism in which the inflaton is identified as one of the real parts of the moduli. 
Although in Ref.~\cite{Abe:2014vca}, both the small- and large-field inflation are discussed in the light of recent Planck results, in this paper, we restrict ourselves to the small-field inflation, for simplicity.\footnote{The extension to the large-field inflation is straightforward.} 
The inflaton potential is generated by the K\"ahler and superpotential of the pair $(T^3,H_3)$ in Eqs.~(\ref{eq:Kmo}) and (\ref{eq:Wmo}), 
\begin{align}
K&= -\ln {\cal N} + Z_{H_3}(\text{Re}T^3) |H_3|^2, 
\nonumber\\
W_{\rm inf} &=\left( J_{0}^{(3)}-J_{L}^{(3)}e^{-c_3^{(3)}T^{3}} 
\right) H_3,
\label{eq:KWinf}
\end{align}
where 
\begin{align}
Z_{H_3}(\text{Re}T^3) =\frac{1-e^{-2c_3^{(3)}{\rm Re}\,T^{3}}}{c_3^{(3)}{\rm Re}\,T^{3}},
\end{align}
and the effective scalar potential is obtained from Eq.~(\ref{eq:scalarpo}) with the above K\"ahler and superpotential~(\ref{eq:KWinf}), 
\begin{align}
V_{\rm inf} &= e^K K^{H_3\bar{H}_3} |W_{H_3}|^2 \simeq 
\frac{|J_{0}^{(3)} -J_{L}^{(3)}e^{-c_3^{(3)} T^3}|^2}{\langle{\rm Re}\,T^1
\rangle \langle{\rm Re}\,T^2
\rangle (1-e^{-c_3^{(3)}T^3})},
\label{eq:infpo}
\end{align}
where ${\rm Re}\,T^3$ is identified as the inflaton. 
Here, it is supposed that the other moduli $T^{I'}$, stabilizer fields $H_i$ with $I',i=1,2$ are heavier than the pair ($T^3, H_3$) and fixed at their minima. 
This is because the minima of them are fixed by their own superpotential in Eq.~(\ref{eq:refmoduli}), and they can be decoupled from the pair $(T^3,H_3)$ by choosing the parameters in the superpotential~(\ref{eq:Wmo}),
\begin{align}
&J_{0}^{(1)}=J_{0}^{(2)}=\frac{1}{9},\,\,
J_{L}^{(1)}=J_{L}^{(2)}=1,
\label{eq:para1}
\end{align}
and the nonvanishing $U(1)_{1,2,3}$ charges 
of $H_{1,2}$,
\begin{align}
&c_1^{(1)}=c_2^{(2)}=\frac{1}{50}, 
\label{eq:para2}
\end{align}
whereas the constants $J_{0,L}^{(3)}$ are chosen to be small compared with $J_{0,L}^{(1),(2)}$ as shown later. 
Furthermore, in the following analysis, we omit the fluctuation of $H_3$ and $X$, because their minima are fixed around the origin by the Hubble-induced mass during the inflation. Im\,$T^3$ is also fixed at the origin during and after the inflation. 
They can be checked that the fluctuations of these fields are negligible to the inflaton dynamics as explicitly shown in Appendix.~\ref{app:oscinf}. 

\begin{figure}[t]
\centering \leavevmode
\includegraphics[width=90mm]{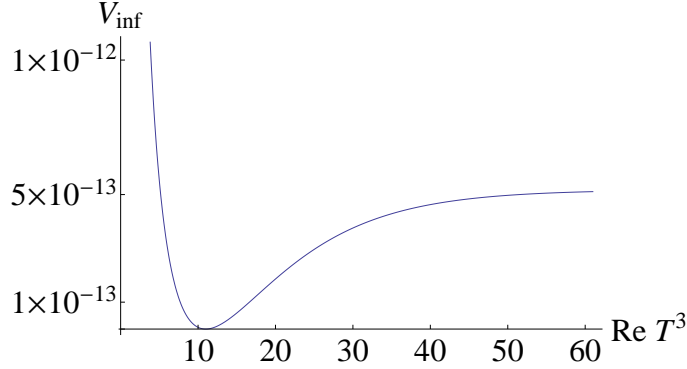}
\caption{Inflaton potential $V_{\rm inf}$ on the ${\rm Im}\,T^1=0$ hypersurface.}
\label{Inflatonpotential}
\end{figure}

When Re\,$T^3$ is identified as the inflaton, the effective scalar potential~(\ref{eq:infpo}) is similar to the one in the Starobinski model~\cite{Starobinsky:1980te} and is drawn in Fig.~\ref{Inflatonpotential} with the 
parameters given by Eqs.~(\ref{eq:para1}), (\ref{eq:para2}) and (\ref{eq:para3}). 
From Fig.~\ref{Inflatonpotential}, the inflaton, Re\,$T^3$ can roll its potential slowly down to its minimum from the large value of Re\,$T^3$. 
In order to evaluate the cosmological observables for the cosmic microwave background (CMB) observed by Planck, we define the slow-roll parameters for the inflaton, $\sigma \equiv \text{Re}T^3$,
\begin{align}
\epsilon &\equiv \frac{M_{\rm Pl}^2}{2} \frac{\nabla_\sigma 
V_{\rm inf} K^{\sigma\sigma} \nabla_\sigma V_{\rm inf}}
{V_{\rm inf}^2},
\nonumber\\
\eta &\equiv \frac{\nabla^\sigma\nabla_\sigma V_{\rm inf}}{V_{\rm inf}},
\end{align}
where $\nabla_\sigma$ is the K\"ahler covariant derivative for the field $\sigma$. 
With these slow-roll approximations, the power spectrum of the scalar curvature perturbation, its spectral index, and tensor-to-scalar ratio can be expressed as  
\begin{align}
P_\xi (k)&=\frac{1}{24\pi^2} \frac{V_{\rm inf}}{\epsilon}, \nonumber\\
n_s &= 1+ \frac{d\text{ln} P_{\xi}(k)}{d \ln k} \simeq 1-6\epsilon +2\eta,
\nonumber\\
r &=16\epsilon. 
\end{align}
The recent data reported by the Planck Collaboration shows the almost scale invariant spectrum and the upper limit of $r$~\cite{Ade:2015lrj},
\begin{align}
P_\xi (k)\simeq 2.20\pm 0.10 \times 10^{-9},\,\,\,
n_s = 0.9655\pm 0.0062,\,\,\,
r < 0.11, 
\end{align}
at the scale $k_\ast =0.05\,[\text{Mpc}^{-1}]$. 
The inflaton dynamics is obeyed by its equation of motion,
  \begin{equation}
\sigma^{''} = -\left(1-\frac{g_{\sigma\sigma} (\sigma^{'})^2}{6} \right)\left( 3\sigma^{'} +6\sigma^2 \frac{V^{'}}{V}\right) +\frac{ (\sigma^{'})^2}{\sigma^{'}},
\label{eq:eom}
\end{equation}
where $'$ denotes the $d/dN$ by employing the number of $e$-foldings $N$ rather than time;
  \begin{equation}
a(t) =e^N,\;\;\; \frac{d}{dt} =\frac{dN}{dt}\frac{d}{dN} =H\frac{d}{dN},
  \end{equation}
where $a(t)$ is the scale factor of $4$D spacetime. 
The metric $g_{\sigma\sigma}$ is connected to the K\"{a}hler metric $K_{T^3\bar{T}^3}$ such that $\frac{1}{2}g_{\sigma\sigma} \partial \sigma \partial \sigma  =K_{T^3\bar{T}^3} \partial T^3 \partial \bar{T}^3$ and $\Gamma^\sigma_{\sigma\sigma} =-1/\sigma$ is the Christoffel symbol. 
As a result, it is found that the power spectrum of scalar curvature perturbation, its spectral index and tensor-to-scalar ratio are consistent with the current cosmological data,
\begin{align}
P_\xi (k)\simeq 2.2\times 10^{-9},\,\,\,
n_s \simeq 0.96,\,\,\,
r \simeq 10^{-5},
\end{align}
with the enough $e$-foldings $N\simeq 58$ and then the parameters are chosen as,
\begin{align}
J_{0}^{(3)}=\frac{1}{4000},\,\,J_{L}^{(3)}=\frac{3}{4000},\,\,
c_3^{(3)}=\frac{1}{10}. 
\label{eq:para3}
\end{align}
The running of the scalar spectral index is negligible, relative to the current observational sensitivity. 

When the numerical values of the parameters are chosen as those in Eqs.~(\ref{eq:para1}), (\ref{eq:para2}) and (\ref{eq:para3}), the moduli VEVs become
\begin{equation}
(\text{Re} \langle T^1\rangle, \text{Re} \langle T^2\rangle, \text{Re} \langle T^3\rangle ) \simeq (110,\;110,\;11),
\label{eq:movev}
 \end{equation}
in the unit $M_{\rm Pl}=1$. According to these moduli VEVs, the typical Kaluza-Klein mass scale is found as 
\begin{equation}
M_{c} =\frac{\pi}{L} \simeq \frac{\pi}{\langle {\cal N}^{1/2}\rangle} \simeq 2.1\times 10^{16} \text{GeV},
\end{equation}
which is close to the GUT scale due to the mild large volume of the fifth dimension, $\langle {\cal N}^{1/2}\rangle \simeq 364$. 
The mass of moduli $T^{I'}$ and stabilizer fields $H_i$ are also given by 
\begin{align}
m_{T^1}\simeq m_{T^2}\simeq m_{H_1}\simeq m_{H_2}\simeq 
4.8\times 10^{15}\,{\rm GeV},\,\,\,
m_{T^3}\simeq m_{H_3}\simeq 4\times 10^{12}\,{\rm GeV},
\label{eq:momass}
\end{align}
and their {\it F}-terms become 
\begin{align}
&
F^{T^1} \simeq F^{T^2}\simeq F^{H_1}\simeq F^{H_2} \simeq 1\times 10^{-42},
F^{T^3}\simeq F^{H_3}\simeq 1.6\times 10^{-36},
\label{eq:moFterm}
\end{align}
in the unit $M_{\rm Pl}=1$. 

So far, we have specified the parameters relevant for the moduli and stabilizer fields. 
The parameters in the K\"ahler and superpotential~(\ref{eq:KWX}) for the SUSY-breaking sector are considered as 
\begin{align}
\nu\simeq -1.567\cdots \times 10^{-14},\,\,
w=-6\times 10^{-14},\,\,
\Lambda= 10^{-4},
c_{X}^{(1)}=\frac{3}{10},\,\, 
c_{X}^{(2)}=\frac{1}{10},\,\,
c_X^{(3)}=0,
\label{eq:para4}
\end{align}
where $\nu$ is proper chosen as realizing the Minkowski minimum. 
Then, the mass of the gravitino and the mass and {\it F}-term of $X$ are obtained as
\begin{align}
&m_{3/2}\simeq 395\,{\rm GeV},\,\,\,m_X\simeq 6\times 10^{8}\,{\rm GeV},\,\,\,
\frac{F^X}{M_{\rm Pl}} \simeq 4541\,{\rm GeV},
\label{eq:gramass}
\end{align}
which implies that the gravitino mass is suppressed by the K\"ahler metric of the SUSY-breaking field, $K_{X\bar{X}}\simeq 1/(c_X\cdot {\rm Re}T) \simeq 0.023$ as discussed in Sec.~\ref{subsec:gravitino} and the concrete sparticle spectra are shown in Sec.~\ref{subsec:Higgsmass}.

\subsection{Moduli-induced gravitino problem and reheating temperature}
\label{subsec:gravitino_problem}
As mentioned in Sec.~\ref{subsec:moduli_stabilization}, the moduli 
and stabilizer fields are so heavy that they decay into 
the particles in the MSSM before the BBN. 
However, even if they are much heavier than O($100$ TeV), 
it has to be taken into account of the cosmological problem, e.g., 
the moduli-induced gravitino problem~\cite{Endo:2006zj,Nakamura:2006uc}. 

The moduli decay width into the gravitino 
pair can be evaluated by the couplings between moduli and gravitinos in the unitary gauge,
\begin{align}
{\cal L}_{3/2} = \epsilon^{\mu\nu\rho\sigma} {\bar \psi}_\mu {\bar \sigma}_\nu \tilde{D}_\rho \psi_\sigma -e^{K/2} W^\ast \psi_\mu \sigma^{\mu\nu} \psi_\nu -e^{K/2} W {\bar \psi}_\mu {\bar \sigma}^{\mu\nu} {\bar \psi}_\nu, 
\label{unitarygauge}
\end{align}
where  $\psi_\mu$ denotes the gravitino in two-component formalism 
and the relevant covariant derivatives of the gravitino are 
$\tilde{D}_\rho \psi_\sigma =\partial_\rho \psi_\sigma +\frac{1}{4}(K_J \partial_\rho \phi^J -K_{\bar{J}}\partial_\rho {\bar \phi}^{{\bar J}})$. 
After carrying out a field-dependent chiral transformation,
\begin{gather}
\psi_\mu \rightarrow \left( \cfrac{W}{{\overline W}}\right)^{-1/4} \psi_\mu,
\end{gather}
the Lagrangian~(\ref{unitarygauge}) is simplified as 
\begin{align}
{\cal L}_{3/2} &= \epsilon^{\mu\nu\rho\sigma} {\bar \psi}_\mu {\bar \sigma}_\nu \partial_\rho \psi_\sigma +\cfrac{\epsilon^{\mu\nu\rho\sigma}}{4} (G_J \partial_\rho \phi^J -G_{\bar J} \partial_\rho {\bar \phi}^{\bar J} ){\bar \psi}_\mu {\bar \sigma}_\nu \psi_\sigma -e^{G/2}( \psi_\mu \sigma^{\mu\nu} \psi_\nu +{\bar \psi}_\mu {\bar \sigma}^{\mu\nu} {\bar \psi}_\nu), 
\end{align}
where $G=K +\text{ln} |W|^2$ and $G_J=\partial_J G$. 
When we expand the moduli $T^{I'}$ around the vacuum 
given by employing the reference point 
method~(\ref{eq:refmoduli}) and (\ref{eq:variation}), the 
Lagrangian~(\ref{unitarygauge}) reduces to 
\begin{align}
{\cal L}_{3/2} &= -\cfrac{\epsilon^{\mu\nu\rho\sigma}}{2} {\bar \Psi}_\mu \gamma_5 \gamma_\nu \partial_\rho \Psi_\sigma +\cfrac{\epsilon^{\mu\nu\rho\sigma}}{8} (\langle G_{T^{J'}}\rangle \partial_\rho \delta T^{J'} -\langle G_{\overline{T}^{J'}}\rangle \partial_\rho \delta\overline{T}^{{\bar J}'}){\bar \Psi}_\mu \gamma_\nu \Psi_\sigma \nonumber\\
&\hspace{13pt}-\cfrac{1}{4}\langle e^{G/2}\rangle {\bar \Psi}_\mu [\gamma^\mu, \gamma^\nu] \Psi_\nu -\cfrac{1}{8}\langle e^{G/2}\rangle (\langle G_{T^{J'}}\rangle\delta T^{J'} +\langle G_{{\overline T}^{{\bar J}'}}\rangle\delta \overline{T}^{{\bar J}'}) {\bar \Psi}_\mu [\gamma^\mu, \gamma^\nu] \Psi_\nu,   
\end{align}
in the four-component formalism of the gravitino $\Psi_\mu$. 
As shown in Eq.~(\ref{eq:momass}), the moduli and stabilizer fields, 
except for the pair ($T^3,H_3$), are decoupled from the inflaton 
dynamics due to their heavy masses. Therefore, their decays can be 
neglected and do not give the sizable effects in the thermal history of 
the universe. 
In this respect, we focus on the decay processes of $T^3$, 
$H_3$, and SUSY-breaking field $X$. 

\subsubsection{The inflaton decay}
First, we concentrate on the inflaton decay into gravitino pair. 
Since the gravitino wavefunction is described in terms of helicity $\pm 1/2$ components of the gravitino at a high-energy limit by the equivalence theorem, 
the inflaton decay width into the gravitino pair is estimated as
\begin{align}
\Gamma (\sigma^3 \rightarrow \Psi_{3/2}\Psi_{3/2})&\simeq \frac{1}{288\pi \langle K_{T^3 \bar{T}^3}\rangle} \left|\left\langle \frac{D_{T^3}W}{W} \right\rangle \right|^2 \cfrac{m_{T^3}^5}{m_{3/2}^2 
M_{\rm Pl}^2} \simeq \frac{1}{288\pi \langle K_{T^3 \bar{T}^3}
{\cal N}\rangle} 
\cfrac{m_{T^3}m_{3/2}^2}{M_{\rm Pl}^2 } \nonumber\\
&\simeq  1.6\times 10^{-18} \;\;\text{GeV},
\end{align}
by employing the {\it F}-term of modulus~(\ref{eq:momasseana}) and numerical values of the mass, {\it F}-term of inflaton, and gravitino mass given by Eqs.~(\ref{eq:momass}) and (\ref{eq:moFterm}). 
Here, the reduced Planck mass has been explicitly written. 
When the inflaton has the sizable {\it F}-term at the vacuum, the enhancement factor $m_{3/2}^{-2}$, as the longitudinal mode of the gravitino, induces the significant amount of gravitinos which would threaten to destroy the success of BBN. 
However, in our moduli inflation, this direct decay is so suppressed due to the almost vanishing {\it F}-term of the inflaton. 
Therefore, the dominant decay process of inflaton comes from the interactions with the gauge bosons, 
\begin{align}
{\cal L}_{Tgg} 
&= -\cfrac{1}{4 (g_a)^2}F^a_{\mu\nu}F^{a\mu\nu} -\cfrac{1}{4}\xi_{J'}^a\delta T_R^{J'} F^a_{\mu\nu}F^{a\mu\nu} -\cfrac{1}{8}\xi_{J'}^a\delta T_I^{J'} \epsilon^{\mu\nu\rho\sigma} F^a_{\mu\nu}F^a_{\rho\sigma},
\end{align}
where $T_R^{J'}={\rm Re}\,T^{J'}$, $T_I^{J'}={\rm Im}\,T^{J'}$. 
Now, the gauge kinetic functions $f_a(X, T)$ are considered as in Eq.~(\ref{eq:gaugekin}). 
In general, $\xi_X^a X$ in the gauge kinetic function (\ref{eq:gaugekin}) could appear, because the {\it R}-symmetry is explicitly broken by the constant superpotential $w$ in Eq.~(\ref{eq:KWX}). 

The inflaton decay width into the gauge bosons are
\begin{align}
\sum_{a=1}^3\Gamma (\sigma^3 \rightarrow g^{(a)}g^{(a)})&\simeq 
\sum_{a=1}^3\cfrac{N_G^a}{128\pi } \left\langle\cfrac{\xi_3^a}{\sqrt{2K_{T^3\bar{T}^3}}}\right\rangle^2\cfrac{m_{T^3}^3}{M_{\rm Pl}^2} 
\simeq 3.95\,\,\, {\rm GeV},
\end{align}
with the numerical values of mass and VEV of modulus~(\ref{eq:movev}), (\ref{eq:momass}), where $N_G^a$ are the number of the gauge bosons for the gauge groups $U(1)_Y$, $SU(2)_L$, $SU(3)_C$ and the nonvanishing coefficients in the gauge kinetic function are chosen as $\xi_3^1=\xi_3^2=\xi_3^3=0.22$ to realize the gauge coupling unification at the GUT scale $M_{\rm GUT}\simeq 2\times 10^{16}\,{\rm GeV}$. 
Although there are the other decay processes via the inflaton decay into the gauginos $\lambda^a$ given by the interactions,
\begin{align}
{\cal L}_{T\lambda\lambda} &= -\cfrac{i}{2} \sum_a {\rm Re}\,f_a(\lambda^a \sigma^\mu D_\mu \bar{\lambda^a} +({\rm H.c.})) +\cfrac{i}{2} \sum_a {\rm Im}\,f_a D_\mu (\lambda^a \sigma^\mu \bar{\lambda^a}) \nonumber\\ 
& \hspace{13pt}+\sum_a\left(\cfrac{1}{4}\frac{\partial f_a}{\partial T^{I'}} F^{T^{I'}} \lambda^a\lambda^a +({\rm H.c.})\right),  
\end{align}
where $D_\mu$ is the covariant derivative for gaugino, such decay channels are suppressed by the small masses of gauginos and almost vanishing {\it F}-term of inflaton such as
\begin{align}
&\sum_{a=1}^3\Gamma  (\sigma^3 \rightarrow \tilde{\lambda^a}\tilde{\lambda^a}) \simeq 
\sum_{a=1}^3\frac{m_{T^3}}{16\pi} \frac{(\xi_3^a)^2 m_{\lambda^a}^2}{M_{\rm Pl}^2} \simeq 1.5\times 10^{-21}\,\text{GeV},
\end{align}
with $m_{\lambda^3}\simeq 1.5\,{\rm TeV}$ and the derivative of {\it F}-term for the inflaton, 
\begin{align}
&\left\langle\frac{\partial F^{T^3}}{\partial T^3}\right\rangle 
=\left\langle\frac{\partial}{\partial T^3} e^{K/2} 
\left( K^{T^3\bar{T}^3} |D_{T^3} W|^2 +K^{T^3\bar{H}^3} D_{T^3} W \overline{D_{H^3} W}\right) \right\rangle \sim O\left( \frac{m_{3/2}^4}{m_{T^3}^2} \right).
\end{align}

The decays from the inflaton into sfermions are also suppressed because of the factor, $m_{\text{sfermion}}/m_{T^3}$, if the masses of sfermions are of O($1$~TeV). 
Other decays from the inflaton into the fermion pairs and quark-quark-gluon are negligible due to their small masses and phase factors, respectively, as pointed out in Ref.~\cite{Nakamura:2006uc}. 
The $\mu$-term does not give the sizable effects for the inflaton decay process, because we consider the tiny $\mu$-term ($\sim$ $500$~GeV) in the light of naturalness as shown in Sec.~\ref{subsec:Yukawa}. 
Finally, we comment on a single gravitino production via the inflaton decay into the modulino and gravitino. 
Since the mixing terms between $T^3$ and $H_3$ in the mass squared matrices are controlled by the SUSY-breaking scale, i.e., the gravitino mass, the mass 
difference between the inflaton and modulino as its superpartner is of the order of the gravitino mass. 
Therefore, the inflaton decay width into the modulino $\tilde{\sigma}^3$ and gravitino is suppressed by the phase factor $m_{3/2}/m_{T^3}$, 
\begin{align}
\Gamma (\sigma^3 \rightarrow {\tilde{\sigma}^3}\Psi_\text{3/2}) &\simeq \frac{1}{48\pi} \left(\frac{m_{T^3}}{M_{\rm Pl}}\right)^2 \left(\frac{m_{3/2}}{m_{T^3}}\right) m_{3/2} 
\simeq 7.2\times 10^{-22}\,\text{GeV},
\end{align}
with $m_{3/2}=395\,{\rm GeV}$, $m_{T^3}=4\times 10^{12}\,{\rm GeV}$ given by Eqs.~(\ref{eq:momass}) and (\ref{eq:gramass}). 
The inflaton decay into the SUSY-breaking field $X$ is also suppressed, because there is no tree-level interaction between $X$ and $T^3$ due to the vanishing $U(1)_3$ charge of $X$. 
As a result, the branching ratios of the moduli decaying into the gravitino(s) are summarized as 
\begin{align}
&\Gamma_{\text{all}}^{\sigma^3} \equiv 
\Gamma (\sigma^3 \rightarrow \text{all}) 
\simeq \sum_{a=1}^3\Gamma (\sigma^3 \rightarrow g^{(a)}g^{(a)}) 
\simeq 3.95\,\,\, {\rm GeV}, \nonumber \\
&\text{Br}(\sigma^3 \rightarrow \Psi_\text{3/2} \Psi_\text{3/2}) 
\simeq \frac{\Gamma (\sigma^3 \rightarrow \Psi_\text{3/2} \Psi_\text{3/2})}{\Gamma_{\text{all}}^{\sigma^3}} \simeq 1.4\times 10^{-20}, \nonumber \\
&\text{Br}(\sigma^3 \rightarrow \tilde{\sigma}^3 \Psi_\text{3/2}) \simeq \frac{\Gamma (\sigma^3 \rightarrow \tilde{\sigma}^3 \Psi_\text{3/2}) }{\Gamma_{\text{all}}^{\sigma^3}} \simeq 1.8\times 10^{-22},
\label{eq:totdec}
\end{align}
and then the reheating temperature is roughly estimated by equaling the expansion rate of the Universe to the total decay width of inflaton, 
\begin{align}
\Gamma_{\rm all}^{\sigma^3} = H_R 
\Leftrightarrow T_{R} = \left( \cfrac{\pi^2 g_\ast}{90}\right)^{-1/4} \sqrt{\Gamma_{\rm all} M_{\rm Pl}} \simeq  1.38\times 10^9~{\rm GeV},
\label{eq:totreh}
\end{align}
where $H_R=H(T_R)$ and $g_\ast (T_R) =915/4$ is the effective degrees of freedom of the radiation in the MSSM at the reheating.  
The gravitino yield $Y_{3/2}$ via the inflaton decay is suppressed due to the tiny branching ratio of the inflaton decay into the gravitino(s),
\begin{equation}
Y_{3/2} =\frac{n_{3/2}}{s} \simeq \text{Br}(\sigma^3 \rightarrow \tilde{\sigma}^3 \Psi_\text{3/2}) \frac{3T_R}{4m_{T^3}} \simeq 3.8\times 10^{-24},
\end{equation} 
with $m_{3/2}=395\,{\rm GeV}$, $T_R=1.38\times 10^{9}\,{\rm GeV}$ $s=4\rho/3T$, where $n_{3/2}$, $s$, and $\rho$ are the number density of the gravitino, entropy, and energy density of the Universe, respectively. 
Now it is supposed that the coherent oscillation of the inflaton field dominates the energy density of the Universe after the inflation and there is no entropy production after the inflation as shown later. 

It is remarkable that the supersymmetric moduli stabilization is important to suppress the direct decays from the inflaton into the gravitino(s) which give the solution to the cosmological moduli problem, especially the moduli-induced gravitino problem. 
The other gravitino production from the stabilizer fields, the SUSY-breaking field, and the thermal bath can be estimated in the next section.

\subsubsection{The decay of stabilizer and SUSY-breaking 
fields}
The stabilizer field $H_3$ is stabilized at the origin during the inflation and after that, Re\,$H_3$ oscillates around its vacuum~(\ref{eq:variation}) deviated from the supersymmetric one~(\ref{eq:refmoduli}). 
On the other hand, Im\,$H_3$ and Im\,$X$ evolve to the origin during inflation and do not oscillate after the inflation as shown in Appendices.~\ref{app:oscvac} and \ref{app:oscinf}. 
Similarly, the SUSY-breaking field Re\,$X$ oscillates around its vacuum after the inflation. 
From the analyses in Appendices.~\ref{app:oscvac} and \ref{app:oscinf}, the amplitudes of both fields are found as 
\begin{align}
\Delta h_3 \simeq \frac{m_{3/2}}{m_{H_3}},\,\,\,\,\,
\Delta x \simeq \left(\frac{m_{3/2}}{m_{X}}\right)^2,
\end{align}
with $h_3={\rm Re}\,H_3$ and $x={\rm Re}\,X$. 
By comparing their masses given in Eq.~(\ref{eq:momass}) with the reheating temperature (\ref{eq:totreh}), the coherent oscillations of both fields $h_3$ and $x$ start before the reheating process. 
When $H_3$ does not couple to the fields in the MSSM, the dominant decay process is
\begin{align}
\Gamma_{\rm all}^{h_3}\equiv 
\Gamma (h_3 \rightarrow \Psi_{3/2}\Psi_{3/2})&\simeq \frac{1}{288\pi K_{H_3 \bar{H}_3} } \left|\left\langle \frac{D_{H_3}W}{W} \right\rangle \right|^2 \cfrac{m_{H_3}^5}{m_{3/2}^2 M_{Pl.}^2} \simeq \frac{1}{288\pi \langle K_{H_3 \bar{H}_3}{\cal N}\rangle}
\cfrac{m_{H_3}^3}{M_{\rm Pl}^2} \nonumber\\
&\simeq  0.02 \;\;\text{GeV},
\end{align}
which implies the decay time of $h_3$ is smaller than the time of the coherent oscillation of $h_3$ and reheating, that is, $H^{h_3}_{\rm osc}>H(T_R)>H^{h_3}_{\rm dec}$, with $H^{h_3}_{\rm osc}\simeq m_{h_3}$ and 
$H^{h_3}_{\rm dec}\simeq \Gamma (h_3 \rightarrow \Psi_{3/2}\Psi_{3/2})$. 
Here and in what follows, $H_{\rm R}$, $H_{\rm osc}^{\Phi}$, and $H_{\rm dec}^{\Phi}$ refer to the Hubble parameters at the time of reheating, beginning of oscillation of relevant fields $\Phi$, 
and decay of $\Phi$. The scale factors of $4$D spacetime $a_R$, $a_{\rm osc}^{\Phi}$, and $a_{\rm dec}^{\Phi}$ are also defined in the same way as the Hubble parameters, $H_{\rm R}$, $H_{\rm osc}^{\Phi}$, and $H_{\rm dec}^{\Phi}$. 
The energy density of coherent oscillation $h_3$ is 
\begin{align}
\rho_{h_3} \simeq \frac{1}{2}m_{H_3}^2 (\Delta h_3)^2 
\simeq \frac{1}{2}m_{3/2}^2
\left( \frac{a}{a^{h_3}_{\rm osc}}\right)^{-3},
\end{align}
where $a^{h_3}_{\rm osc}$ stands for the scale factor at the time when $h_3$ begins to oscillate and $\rho_{h_3}$ is converted into the gravitino yield hereafter,
\begin{align}
Y_{3/2}^{h_3} =\frac{2 \rho_{h_3}}{m_{H_3}s}\simeq 
 \frac{1}{4}\frac{m_{3/2}^2 T_R}{m_{H_3}^3} \simeq 
8.2\times 10^{-25},
\end{align}
with $m_{3/2}=395\,{\rm GeV}$, $T_R=1.38\times 10^{9}\,{\rm GeV}$, and $m_{H_3}=4\times 10^{12}\,{\rm GeV}$.
Here we employed that the entropy production from $h_3$ can be neglected. 
In our model, the following inequality is satisfied due to the tiny mass of the gravitino and then $h_3$ does not dominate the Universe and 
release the significant entropy,
\begin{align}
1\gg \frac{\rho_{h_3}}{\rho_R}\biggl|_{T=T_{\rm dec}^{h_3}}
=\frac{\rho_{h_3}}{\rho}\biggl|_{\rm end} 
\left( \frac{T_R}{T_{t_{\rm dec}^{h_3}}}\right) 
\simeq \frac{m_{3/2}^2M_{\rm Pl}^2}{2V_{\rm inf}}
\left( \frac{T_R}{T_{t_{\rm dec}^{h_3}}}\right) ,
\end{align}
where $\rho_{h_3}$, $\rho_R$ are the energy densities of $h_3$ and radiation, respectively. 
$\rho|_{\rm end}=V_{\rm inf}\simeq {\cal O}(10^{-13})$  denotes the energy density at the end of inflation and 
$T_{\rm dec}^{h_3}$ is the decay temperature of $h_3$,
\begin{align}
T_{\rm dec}^{h_3}= \left( \cfrac{\pi^2 g_\ast}{90}\right)^{-1/4} \sqrt{\Gamma_{\rm all}^{h_3}
M_{\rm Pl}} \simeq  8.6\times 10^7~{\rm GeV}.
\label{eq:dech3}
\end{align}

Furthermore, the SUSY-breaking field also produces the gravitinos through the following dominant decay channel:
\begin{align}
\Gamma (x \rightarrow \Psi_{3/2}\Psi_{3/2})& 
\simeq \frac{1}{288\pi \langle K_{X \bar{X}}\rangle}
\left|\left\langle \frac{D_{X}W}{W} \right\rangle \right|^2
\cfrac{m_{X}^5}{m_{3/2}^2M_{\rm Pl}^2}
\simeq \frac{1}{288\pi \langle K_{X \bar{X}}\rangle}
\left|\frac{\nu}{w} \right|^2
\cfrac{m_{X}^5}{m_{3/2}^2M_{\rm Pl}^2}.
\end{align}
With the parameters~(\ref{eq:para4}), the VEVs of moduli (\ref{eq:movev}), and mass of $\Psi_{3/2}$ and $X$ (\ref{eq:gramass}), the total decay width of $X$ then becomes
\begin{align}
\Gamma_{\rm all}^x\equiv 
\Gamma (x \rightarrow \Psi_{3/2}\Psi_{3/2})
\simeq  3.7\times 10^{8} \;\;\text{GeV}.
\label{eq:totdecx}
\end{align}
Therefore, the decay time of $x$ is smaller than that of reheating, that is, $H^x_{\rm osc}>H^x_{\rm dec}\gg H(T_R)$, with $H^x_{\rm osc}\simeq m_X$ and $H^x_{\rm dec}\simeq \Gamma (x \rightarrow \Psi_{3/2}\Psi_{3/2})$. 
The energy density of the coherent oscillation $x$ is converted into that of the gravitino as 
\begin{align}
\rho_{x} \simeq \frac{1}{2}m_{x}^2 (\Delta x)^2 
\simeq \frac{1}{2}\left( \frac{m_{3/2}^4}{m_x^2}\right) 
\left( \frac{a^x_{\rm dec}}{a^x_{\rm osc}}\right)^{-3}
\left( \frac{a_R}{a^x_{\rm dec}}\right)^{-4},
\end{align}
at the time of reheating, where the gravitino is relativistic at the time of production. 
By employing the scale factors, 
\begin{align}
&\frac{a_R}{a_{\rm osc}^{\sigma^3}}=
\left( \frac{\sqrt{6}\Gamma_{\rm all}^{\sigma^3}}{m_{T^3}}\right)^{-2/3},
\,\,\,
\frac{a_{\rm osc}^x}{a_{\rm osc}^{\sigma^3}}=
\left( \frac{\sqrt{6}m_X}{m_{T^3}}\right)^{-2/3},
\,\,\,
\frac{a_{\rm dec}^x}{a_{\rm osc}^{\sigma^3}}=
\left( \frac{6(\Gamma_{\rm all}^x\,m_X)^2}{m_{3/2}^4}\right)^{-1/3},
\end{align}
the gravitino yield is 
\begin{align}
Y_{3/2}^{x} =\frac{2 \rho_{x}}{m_{x}s}\simeq 
\frac{3}{2}\frac{T_R}{m_X}
\left(\frac{m_{3/2}}{m_X}\right)^{16/3}
\left( \frac{\Gamma_{\rm all}^{\sigma^3}}
{\Gamma_{\rm all}^{x}}\right)^{2/3}
\simeq 
2\times 10^{-32},
\end{align}
with the numerical values given by Eqs.~(\ref{eq:gramass}), (\ref{eq:totdec}), (\ref{eq:totreh}), and (\ref{eq:totdecx}). 
It is found that the gravitino production via $x$ decay is suppressed by the tiny mass of the gravitino and it is not the dominant source for the relic abundance of the gravitino. 
The entropy production from $x$ can be also neglected in the same way as that of $h_3$. 
As pointed out in Ref.~\cite{Nakayama:2012hy}, under $m_{3/2}\ll m_{\rm X}\ll m_{T^3}\leq \Lambda$, the gravitino production is significantly relaxed and this condition is satisfied in our model. 
Note that when $\Lambda$ is smaller than the inflaton mass, we have to take account of the inflaton decay into the fields in the hidden sector.

\section{Gravitino dark matter and the Higgs boson mass}
\label{sec:gravitino}
\subsection{Yukawa couplings and naturalness}
\label{subsec:Yukawa}
Before estimating the relic abundance of the gravitino, we specify the Yukawa couplings and $\mu$-term in the superpotential which can be only introduced at the orbifold fixed points $y=0,L$, 
where the SUSY is reduced to $4$D ${\cal N}=1$. 
As stated in Sec.~\ref{subsec:4Deff}, we consider the Yukawa interactions in the MSSM at the orbifold fixed point $y=0$,
\begin{equation}
W_{\rm Yukawa} = \lambda_{ij}^u{\cal Q}_i{\cal H}_u{\cal U}_j
 +\lambda_{ij}^d{\cal Q}_i{\cal H}_d{\cal D}_j +\lambda_{ij}^e{\cal L}_i{\cal H}_d{\cal E}_j +\lambda_{ij}^n{\cal L}_i{\cal H}_u N_j, 
\label{eq:wmssm}
\end{equation}
where $\lambda_{ij}^{u, d, e, n}$ are the holomorphic Yukawa coupling constants and are supposed to be of ${\cal O}(1)$. 
After the canonical normalization of fields in the MSSM, the physical Yukawa couplings are expressed as 
\begin{align}
 y_{ij}^u &= \frac{\lambda_{ij}^u}
 {\sqrt{\langle Y_{{\cal Q}_i}Y_{{\cal H}_u}Y_{{\cal U}_j} \rangle}}, \;\;\;\;
 y_{ij}^d = \frac{\lambda_{ij}^d}
  {\sqrt{\langle Y_{{\cal Q}_i}Y_{{\cal H}_d}Y_{{\cal D}_j} \rangle}}, \;\;\;\;
 y_{ij}^e = \frac{\lambda_{ij}^e}
  {\sqrt{\langle Y_{{\cal L}_i}Y_{{\cal H}_d}Y_{{\cal E}_j} \rangle}}, \;\;\;\;
 y_{ij}^n = \frac{\lambda_{ij}^n}
  {\sqrt{\langle Y_{{\cal L}_i}Y_{{\cal H}_u}Y_{{\cal N}_j} \rangle}}, \nonumber\\
\end{align}
where 
\begin{equation}
Y_a \equiv 2{\cal N}^{1/3}({\rm Re}T)\left\{Y(c_a\cdot T) +\tilde{\Omega}^{(4)}_{a,X}({\rm Re}T)|X|^2 +{\cal O}(|X|^4)\right\}
\end{equation}
The function~$Y(z)$ is always positive, and approximated as 
 \begin{equation}
 Y(z) \equiv \frac{1-e^{-2{\rm Re}z}}{2{\rm Re}z}
\simeq 
\begin{cases}  
\frac{1}{2\text{Re} z},  
& \text{Re} z>0 \\ \frac{1}{2|\text{Re} z|}\text{exp}(2|\text{Re} z|). & \text{Re} z<0. 
\end{cases}
\label{yabc}
 \end{equation}
In the 5D viewpoint, the wavefunctions of fields are localized toward $y=0$ ($y=L$) in the case that $c_a\cdot\langle \text{Re} T\rangle$ is positive (negative). 
As can be seen in Eq.~(\ref{yabc}), $y_{ij}^{u,d,e,n}$ are of ${\cal O}(1)$ or exponentially small when all the relevant fields are localized toward $y=0$ or $y=L$, respectively. 

Therefore, we expect that the mass hierarchies of elementary particles and the extreme smallness of the neutrino masses can be realized even in the case of Dirac neutrinos. 
In fact, when we choose the $U(1)_{I'}$ charges and ${\cal O}(1)$ values of the holomorphic Yukawa couplings $\lambda^{u,d,e,n}_{i,j}$ in Tables~\ref{charge} and~\ref{holomorphic Yukawa couplings}, the observed masses and mixing angles of quarks and leptons at the electroweak scale can be realized. 
Here, we employ the full one-loop RG equations of the MSSM from the GUT to the EW scale. 
It is remarkable that the flavor structure of soft SUSY-breaking terms is determined by the $U(1)_{I'}$ charge assignment as can be seen in the K\"ahler potential~(\ref{eq:effKahler}). 
In fact, the soft SUSY-breaking terms at the GUT scale are determined by the following formula~\cite{Choi:2005ge,Kaplunovsky:1993rd}: 
\begin{align}
 M_a &= \langle F^I\partial_I \text{ln}\left(\text{Re} f^a \right)\rangle, 
\nonumber\\
 m_{Q_\alpha}^2 &= 
 -\langle F^I\bar{F}^{\bar{J}}\partial_I\partial_{\bar{J}} \text{ln} Y_{Q_\alpha}\rangle, \nonumber\\
 A_{ij}^u &= \langle F^I\partial_I \text{ln}\left( Y_{{\cal H}_u}Y_{{\cal Q}_i}Y_{{\cal U}_j}\right)\rangle,\nonumber\\
A_{ij}^d &= \langle F^I\partial_I\text{ln}\left(Y_{{\cal H}_d}Y_{{\cal Q}_i}Y_{{\cal D}_j}\right)\rangle,\nonumber\\
A_{ij}^e &= \langle F^I\partial_I\text{ln}\left(Y_{{\cal H}_d}Y_{{\cal L}_i}Y_{{\cal E}_j}\right)\rangle,\nonumber\\
A_{ij}^n &= \langle F^I\partial_I\text{ln}\left(Y_{{\cal H}_u}Y_{{\cal L}_i}Y_{{\cal N}_j}\right) \rangle, 
\label{eq:softterms} 
\end{align}
where indices~$I$ and $J$ run over all the chiral multiplets. 
Then, the $U(1)_{I'}$ charge assignment in Table~\ref{charge} and the {\it F}-term of the SUSY-breaking field $X$ given by Eq.~(\ref{eq:gramass}) 
give rise to the soft scalar masses and gluino mass in Table~\ref{tab:sparticlegut}. 
By contrast, the {\it A}-terms are almost vanishing due to the tiny {\it F}-terms of moduli. 
Here and hereafter, we parametrize the ratios of gaugino masses at the GUT scale as 
\begin{align}
r_1=\frac{M_1(M_{\rm GUT})}{M_3(M_{\rm GUT})},\,\,\,
r_2=\frac{M_2(M_{\rm GUT})}{M_3(M_{\rm GUT})},
\label{eq:softterms} 
\end{align}
where $M_1(M_{\rm GUT})$, $M_2(M_{\rm GUT})$, and $M_3(M_{\rm GUT})$ are the bino, wino, and gluino masses at the GUT scale, $M_{\rm GUT}\simeq 2\times 10^{16}\,{\rm GeV}$. 
The ratios of gaugino masses are controlled by the parameters $\xi_X^a$ in the gauge kinetic function~(\ref{eq:gaugekin}) without spoiling the gauge coupling unification due to the tiny VEV of $X$.
\begin{table}[h]
\begin{center}
\begin{tabular}{|l|l|l|} \hline
\rule[-2mm]{0mm}{7mm}
$c_{{\cal Q}_i}^{I'=1}=(0.1,~0.1,~1.1)
$ & $c_{{\cal L}_i}^{I'=1}=(0.1,~0.1,~1.6)$ & $c_{{\cal H}_u}^{I'=1}=0
$\\ 
$c_{{\cal Q}_i}^{I'=2}=(-0.1,-0.1, 0.8)$ 
& $c_{{\cal L}_i}^{I'=2}=(-0.1,-0.1,~0)$ & $c_{{\cal H}_u}^{I'=2}=0.1$ \\ 
$c_{{\cal Q}_i}^{I'=3}=(0.1,~0.4,~1)$ 
& $c_{{\cal L}_i}^{I'=3}=(0.1,~0.5,~0)$ & $c_{{\cal H}_u}^{I'=3}=-0.9$ \\ \hline
\rule[-2mm]{0mm}{7mm}
$c_{{\cal U}_i}^{I'=1}=(0.1,~0.1,~0.6)
$ & $c_{{\cal E}_i}^{I'=1}=(0.1,~0.2,~0.2)$ & $c_{{\cal H}_d}^{I'=1}=0
$\\ 
$c_{{\cal U}_i}^{I'=2}=(-0.1,-0.1, 0.3)$ 
& $c_{{\cal E}_i}^{I'=2}=(-0.1,-0.1,0)$ & $c_{{\cal H}_d}^{I'=2}=0$ \\ 
$c_{{\cal U}_i}^{I'=3}=(-0.2,~0.2,~1)$ 
& $c_{{\cal E}_i}^{I'=3}=(-0.2,~0,~-0.5)$ & $c_{{\cal H}_d}^{I'=3}=-0.1$ \\ \hline
\rule[-2mm]{0mm}{7mm}
$c_{{\cal D}_i}^{I'=1}=(0.1,~0.1,~0.2)
$ & $c_{N_i}^{I'=1}=(0.1,~0.1,~0.1)
$ & \\ 
$c_{{\cal D}_i}^{I'=2}=(-0.1,-0.1,0)$ 
& $c_{N_i}^{I'=2}=(-0.3,-0.3,-0.3)$ &   \\ 
$c_{{\cal D}_i}^{I'=3}=(0.3,~0.2,~-0.5)$ & $c_{N_i}^{I'=3}=(-0.7,~-0.7,~-0.7)$ 
&   \\ \hline
\end{tabular}
\caption{$U(1)_{I'}$ flavor charges of the quarks, leptons, and Higgs.}
\label{charge}
\end{center}
\end{table}
\begin{table}[h]
\begin{center}
\begin{tabular}{|c|c|} \hline
$|\lambda^u_{ij}|$ & $|\lambda^d_{ij}|$ \\ \hline
\begin{minipage}{0.38\linewidth}
\begin{eqnarray} 
\left( 
\begin{array}{ccc}
0.32 & 0.35 & 0.95 \\
0.22 & 0.42 & 0.33 \\
0.51 & 0.48 & 1.5 
\end{array}
\right) 
\nonumber
\end{eqnarray} \\*[-20pt]
\end{minipage}
& 
\begin{minipage}{0.38\linewidth}
\begin{eqnarray} 
\left( 
\begin{array}{ccc}
0.45 & 0.5 & 0.59 \\
0.28 & 0.24 & 0.38 \\
1.03 & 1.02 & 0.81 
\end{array}
\right) 
\nonumber
\end{eqnarray} \\*[-20pt]
\end{minipage} 
\\ \hline
$|\lambda^e_{ij}|$ & $|\lambda^n_{ij}|$ \\ \hline
\begin{minipage}{0.38\linewidth}
\begin{eqnarray} 
\left( 
\begin{array}{ccc}
0.28 & 0.22 & 0.52 \\
0.4 & 1.15 & 0.31 \\
0.8 & 1.02 & 1.05 
\end{array}
\right) 
\nonumber
\end{eqnarray} \\*[-20pt]
\end{minipage}
&
\begin{minipage}{0.38\linewidth}
\begin{eqnarray} 
\left( 
\begin{array}{ccc}
0.77 & 0.85 & 0.69 \\
0.25 & 0.98 & 0.58 \\
0.34 & 0.26 & 1.03 
\end{array}
\right) 
\nonumber
\end{eqnarray} \\*[-20pt]
\end{minipage}
\\ \hline
\end{tabular}
\end{center}
\caption{
${\cal O}(1)$ values 
of the holomorphic Yukawa couplings $\lambda^{u,d,e,n}_{ij}$ in the superpotential (\ref{eq:wmssm}).}
\label{holomorphic Yukawa couplings}
\end{table}
\begin{table}[h]
\begin{center}
\begin{tabular}{|c|c||c|c|} \hline 
Sparticles&Mass[GeV] &(S)Particles&Mass[GeV]\\ \hline
$m_{\tilde{{\cal Q}}_1}$ &1682&$m_{\tilde{{\cal L}}_3}$ &2834\\ \hline
$m_{\tilde{{\cal Q}}_2}$ &1530&$m_{\tilde{{\cal E}}_1}$ &1157\\ \hline
$m_{\tilde{{\cal Q}}_3}$ &581&$m_{\tilde{{\cal E}}_2}$ &2390\\ \hline
$m_{\tilde{{\cal U}}_1}$ &1157&$m_{\tilde{{\cal E}}_3}$ &2298\\ \hline
$m_{\tilde{{\cal U}}_2}$ &1698&$m_{\tilde{N}_1}$ &414.5\\ \hline
$m_{\tilde{{\cal U}}_3}$ &799&$m_{\tilde{N}_2}$ &414.5\\ \hline
$m_{\tilde{{\cal D}}_1}$ &1636&$m_{\tilde{N}_3}$ &414.5\\ \hline
$m_{\tilde{{\cal D}}_2}$ &1698 &$M_{{\cal H}_u}$ &1100\\ \hline
$m_{\tilde{{\cal D}}_3}$ &2298 &$M_{{\cal H}_d}$ &298.5\\ \hline
$m_{\tilde{{\cal L}}_1}$ &1682 & $M_{3}$ &550\\ \hline
$m_{\tilde{{\cal L}}_2}$ &1396 & &\\ \hline
 \end{tabular}
 \caption{The soft scalar masses $m_{\tilde{Q}_\alpha}$, the up- and down-type 
Higgs masses $M_{{\cal H}_{u,d}}$, and the gluino mass $M_3$ at the GUT scale. 
The subscripts $\tilde{Q}_\alpha$ denote the mass eigenvalues for the left-handed 
$\tilde{\cal Q}_i$, up-type right-handed $\tilde{\cal U}_i$, down-type right-handed $\tilde{\cal D}_i$ squarks, left-handed $\tilde{\cal L}_i$, right-handed $\tilde{\cal E}_i$ 
charged sleptons, and right-handed sneutrinos $\tilde{N}_i$ with the 
three-generation $i=1,2,3$.}
  \label{tab:sparticlegut}
\end{center}
\end{table} 

On the other hand, the $\mu$-term can be generated by the following superpotential:
\begin{equation}
W_{\mu-{\rm term}} =\sum_{i=1}^3\kappa_i H_i {\cal H}_u{\cal H}_d, 
\end{equation}
where $\kappa_i$ are the $O(1)$ dimensionless couplings, $H_i$ are the stabilizer fields with {\it R}-charge $2$, whereas Higgs chiral superfields do not have the {\it R}-charge. 
These cubic interactions do not affect the moduli stabilization as well as the moduli inflation due to the almost vanishing VEVs of the Higgs fields. 
Thus, it is possible to consider the VEVs of the stabilizer fields $H_i$ as the origin of the $\mu$-term. 
After the canonical normalization of the relevant fields, the $\mu$-term at the GUT scale becomes 
\begin{equation}
\mu =\sum_{i=1}^3 \frac{\kappa_i \langle H_i\rangle}{
\langle Y_{H_i}Y_{{\cal H}_u} Y_{{\cal H}_d} 
\rangle}.
\end{equation}
Especially, in the case of $\kappa_{2}=\kappa_{3}=0$, the scale of the $\mu$-term is chosen as TeV scale,
\begin{align}
\mu \simeq 3.8\times 10^{-3} \frac{m_{3/2}}{m_{H_1}}M_{\rm Pl} 
\simeq {\cal O}(500\,{\rm GeV}),
\end{align} 
where $\kappa_1=2/3$, $m_{3/2}=395\,{\rm GeV}$, $m_{H_1}\simeq 4.8\times 10^{15}\,{\rm GeV}$, and $\langle H_1\rangle \simeq m_{3/2}/m_{H_1}$ are 
given by Eq.~(\ref{eq:variation}) and the factor $3.8\times 10^{-3}$ comes from the mild large volume of the fifth dimension and normalization factors for $H_1$, ${\cal H}_u$, and ${\cal H}_d$. 
The EW symmetry breaking requires the following relation between the mass of the $Z$-boson, $m_Z$ and soft SUSY-breaking masses of the up-type Higgs $m_{{\cal H}_u}$: 
\begin{align}
\frac{m_Z^2}{2} 
&\simeq  -m_{H_u}^2(M_{\rm EW}) -\left| \mu(M_{\rm EW}) \right|^2 
+O\left(\frac{1}{{\rm tan^2}\beta}\right), 
\label{eq:mz}
\end{align}
in the limit of large value of tan$\beta$, where $\mu(M_{\rm EW})$ and $m_{H_u}(M_{\rm EW})$ are the $\mu$-term and $m_{H_u}$ at the EW scale, respectively. 
The VEVs of up- and down-type Higgs fields are denoted by $v_u=v\,{\rm sin}\beta$ and $v_d=v\,{\rm cos}\beta$ with $v=174$\, GeV. 
Thus, the observed $Z$-boson mass indicates $|\mu(M_{\rm EW})| \sim |m_{H_u}(M_{\rm EW})| \sim m_Z$; otherwise $\mu$ and $m_{H_u}$ have to be properly tuned to obtain the EW vacuum. 
We adopt the measure of the degree of tuning the $\mu$-term at the GUT scale as
\begin{equation}
\Delta_\mu = \frac{1}{2}\frac{ \partial \ln m_Z^2}{\partial \ln |\mu|}, 
\label{eq:deltamu}
\end{equation}
and then $100 \times |\Delta_\mu^{-1}|$ \% represents the degree of tuning to obtain the $Z$-boson mass $m_Z=91.2$ GeV~\cite{Barbieri:1987fn}. 
Although the conventional CMSSM scenario requires more severe tuning than the degree of $0.1$ \%, as pointed out in Ref.~\cite{Abe:2007kf}, certain ratios of the nonuniversal gaugino masses at the GUT scale relax the degree of tuning and observed 125 GeV Higgs boson mass at the same time. 

\subsection{Relic abundance of the gravitino}
\label{subsec:Relic}
We are now ready to estimate the relic abundance of the gravitino. 
As stated in Sec. \ref{subsec:gravitino_problem}, there are no significant gravitino productions from the inflaton, moduli, stabilizer, and SUSY-breaking fields after the inflation. 
However, there are two processes to produce the gravitinos associated with the decay of other particles in the MSSM. 

\begin{figure}[t,b]
\centering \leavevmode
\includegraphics[width=80mm]{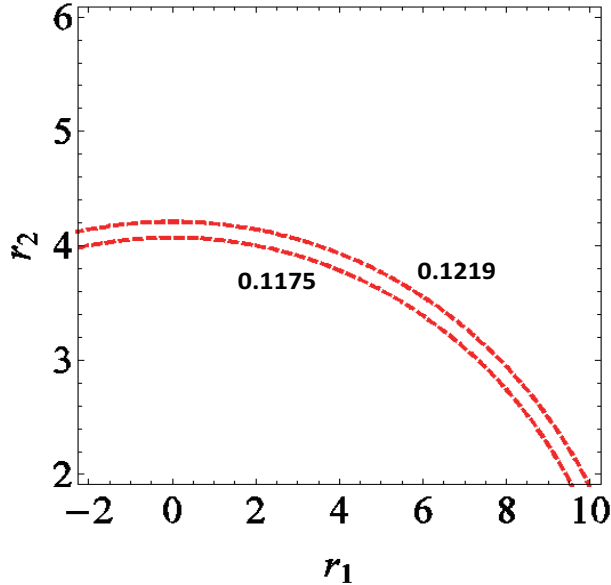}
\caption{Contours of the thermal abundance of the gravitino in the ($r_1,r_2$)-plane.}
\label{fig:thgravitino}
\end{figure}

One of them is the decay from the thermal bath which is constituted of the relativistic particles after the reheating process. 
On the thermal bath, the dominant decay process comes from gauginos into gravitinos, because the couplings between the gravitino and other sparticles are more suppressed than those of gauginos as 
discussed in Refs.~\cite{Bolz:2000fu, Pradler:2006qh}. 
The abundance of the gravitino is estimated as
\begin{equation}
\Omega_{3/2}^{TP} h^2 = \sum_{a=1}^3 \left(1+\cfrac{M_a(T_R)^2 }{3m_{3/2}^2} \right) w_a g_a(T_R)^2~{\rm ln}\left(\cfrac{k_a}{g_a(T_R)} \right)
\left(\cfrac{m_{3/2}}{100 \text{GeV}} \right) \left(\cfrac{T_R}{10^{10} {\rm GeV}} \right),
\label{eq:thermalprod}
\end{equation}
where $w_a$ and $k_a$ are the parameters whose values are defined in Ref.~\cite{Pradler:2006qh} and $h$ is a dimensionless Hubble parameter. 
The thermal production of the gravitino is drawn in Fig.~\ref{fig:thgravitino} in terms of the ratios of gaugino masses at the GUT scale~$M_{\rm GUT}$, $r_1=M_1(M_{\rm GUT})/M_3(M_{\rm GUT})$ and  $r_2=M_2(M_{\rm GUT})/M_3(M_{\rm GUT})$ with $M_{3}(M_{\rm GUT}) =550$ GeV. 
The gaugino masses at the reheating temperature $M_a(T_R)$ can be expressed as $M_a(M_{\rm GUT})$ by employing the one-loop RG equations in the MSSM. 
The Planck Collaboration reported that the abundance of dark matter resides in the range of $0.1175 \le \Omega_{3/2}^{TP} h^2 \le 0.1219$~\cite{Ade:2015lrj}, where the upper and lower limits 
correspond to the dotted curves in Fig.~\ref{fig:thgravitino}. 
Here we assume that the dark matter only consists of the abundance of the thermally produced gravitino. 
\begin{table}[h]
\begin{center}
\begin{tabular}{|c|c|} \hline 
NNLSP(Higgsino-like neutralino) & Mass[GeV]\\ \hline
$\tilde{\chi}_3^0$ &441\\ \hline
NLSPs(right-handed sneutrinos) &mass[GeV]\\ \hline
$\tilde{\nu}_{e_2}$ &415\\ \hline
$\tilde{\nu}_{\mu_2}$ &415\\ \hline
$\tilde{\nu}_{\tau_2}$ &415\\ \hline
LSP(gravitino) &mass[GeV]\\ \hline
$\Psi_{3/2}$ &395\\ \hline
 \end{tabular}\\
 \caption{The masses of NNLSP, NLSPs, and the gravitino 
 at the EW scale for the reference point $(r_1,r_2)=(6,3.5)$. 
The subscripts denote the mass eigenvalues for the sneutrinos 
($\tilde{\nu}$), the Higgsino-like neutralino ($\tilde{\chi}$).}
 \label{tab:LSP}
\end{center}
\end{table} 

The other process is the nonthermal gravitino productions from the NLSP and/or next-to-next-to-lightest supersymmetric particle (NNLSP). 
As shown in Table~\ref{tab:LSP}, when we take the ratios of gaugino masses $(r_1,r_2)=(6,3.5)$ consistent with the observed relic abundance of dark matter in Fig.~\ref{fig:thgravitino}, the NLSPs and NNLSP correspond 
to the degenerated sneutrinos and Higgsino-like neutralino, respectively. 
The relevant sparticle spectra are obtained by employing the one-loop RG equations in the MSSM from GUT to EW scale with $(r_1,r_2)=(6,3.5)$ and the input parameters in Table~\ref{tab:sparticlegut}. 
The full sparticle spectra are shown in the next section. 
Note that the degenerated sneutrinos do not have sizable interactions with the other (s)particles due to the tiny Yukawa couplings of Dirac-type neutrinos and then the soft SUSY-breaking 
masses of right-handed sneutrinos do not receive significant loop corrections. 

Since the gravitino and right-handed sneutrinos are weakly coupled with the other (s)particles, they are not thermalized. 
Thus, the nonthermal gravitino productions from the higgsino-like neutralino and sneutrinos are roughly estimated as 
\begin{equation}
\Omega_{3/2}^{NTP} h^2 = \cfrac{m_{3/2}}{m_{\tilde{\chi}_3^0}} \Omega_{\tilde{\chi}_3^0} h^2, 
\end{equation}
where $m_{\tilde{\chi}_3^0}$ and $\Omega_{\tilde{\chi}_3^0}$ are the mass and the thermal abundance of the Higgsino-like neutralino $\tilde{\chi}_3^0$, respectively. 
The thermal abundance of the Higgsino-like neutralino is known to be small when the $\mu$-term is smaller than wino and bino masses. 
Since the chargino and Higgsino-like neutralino are degenerated,  both decay into the particles of the SM at almost the same decoupled time, 
which leads to the smallness of the thermal abundance of $\tilde{\chi}_3^0$. 
After all, the nonthermal abundance of the gravitino can be neglected,
\begin{equation}
\Omega_{3/2}^{NTP} h^2  \ll 0.11,
\end{equation}
and the total relic abundance of the gravitino is approximated by the thermal abundance of it,\footnote{In this paper, we do not take the gravitino production by the primordial black hole into account~\cite{Khlopov:2004tn}.} 
\begin{equation}
\Omega_{3/2}h^2  \simeq \Omega_{3/2}^{TP}h^2.
\end{equation}

However, the decays of neutralino and sneutrinos into the gravitino dark matter would threaten to spoil the successful BBN. 
The produced right-handed neutrinos via the sneutrino decay into the gravitino are suppressed due to the thermal abundance of $\tilde{\chi}_3^0$, and then they are harmless for the BBN. 
On the other hand, the Higgsino-like neutralino decay into the gravitino affects the BBN. 
The authors of Ref.~\cite{Ishiwata:2007bt} suggest a way to relax the constraints from the BBN by assuming that the NLSP is the Dirac-type right-handed sneutrino. 
Although they consider the bino-like neutralino as the NNLSP, the sparticle spectra are almost the same as our obtained one. 
Because of the small thermal abundance of the Higgsino-like neutralino, it is then expected that our spectra are consistent with the BBN. 

Note that the nonthermal production of the gravitino is enhanced when the bino-like neutralino is NNLSP which corresponds to the small value of $|r_1|$ in Fig.~\ref{fig:thgravitino}. 
In this case, it would break the successful BBN because of the large thermal abundance of the bino-like neutralino~\cite{Kanzaki:2006hm,Feng:2004zu,Ishiwata:2007bt}. 

\subsection{The Higgs boson mass, gravitino dark matter and sparticle spectra}
\label{subsec:Higgsmass}
\begin{figure}[t]
\centering \leavevmode
\includegraphics[width=80mm]{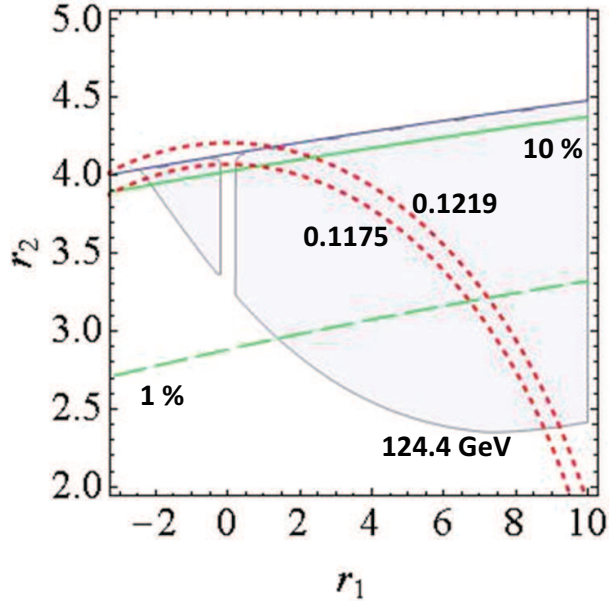}
\caption{The Higgs boson mass, the degree of tuning a $\mu$-term, 
$|\Delta_\mu|\times 100(\%)$, and the relic abundance of the gravitino 
$\Omega_{3/2}h^2$ on the $(r_1,r_2)$-plane. 
In the blue shaded regions, the Higgs boson mass 
resides in the allowed range, $124.4\le m_h\le 126.8\,{\rm GeV}$~\cite{Aad:2012tfa}. 
The green dashed and solid lines show the 1\% and 10\% tuning, respectively. 
The red dashed curves show the relic abundance of the gravitino within 
ranges $0.1179\leq \Omega_{3/2}h^2\leq 0.1215$, reported by 
the Planck Collaboration~\cite{Ade:2015lrj}.}
\label{fig:higgsmugra}
\end{figure}
The ratios of gaugino masses at the GUT scale, $r_1$ and $r_2$, 
are severely constrained by the relic abundance of the gravitino 
as can be seen in Fig.~\ref{fig:thgravitino}.
In this section, we show that the mass of the Higgs boson 
further constrains the ratios of gaugino masses, $r_1$ and 
$r_2$. 
The lightest {\it CP}-even Higgs boson corresponds to the 
SM-like Higgs in the framework of MSSM. 
Without the loop corrections, the Higgs boson mass is 
much lower than the observed mass of the Higgs 
reported by the LHC experiment~\cite{Okada:1990vk}. 
Although, the high-scale SUSY-breaking scenario is a simple 
solution as one of the possibilities to raise the 
Higgs mass, it requires the tuning to obtain the EW vacuum. 
Therefore, we consider the maximal mixing of left- and right-handed 
top squarks to raise the Higgs boson mass without a severe fine-tuning. 

With an approximation that the mass eigenstates of top squarks are nearly degenerate, the mass of the lightest {\it CP}-even Higgs boson cannot be realized. 
Thus, as pointed out in~\cite{Abe:2007kf}, we add the contribution from 
the mass differences between left- and right-handed top squarks in order to 
realize the observed Higgs boson mass and relax the degree of tuning 
at the same time. 
By employing the full one-loop RG equations of the MSSM from 
the GUT to the EW scale, we numerically calculate the Higgs boson 
mass which resides in the range of $124.4 \le m_{h} \le 126.8$~\cite{Aad:2012tfa}, 
which is represented as the blue colored region in 
Fig.~\ref{fig:higgsmugra} and the degree of tuning a 
$\mu$-term, $|\Delta_\mu| \times 100\%$, is also given by 
the green dashed (1\%) and solid lines (10\%), respectively. 
From Fig.~\ref{fig:higgsmugra}, there are the parameter 
spaces which are consistent with the relic abundance of 
the gravitino and the Higgs boson mass reported by the current 
cosmological observations~\cite{Ade:2015lrj} as well as the 
collider experiments~\cite{Aad:2012tfa} without a severe fine-tuning. 

In particular, at the reference point $(r_1, r_2)=(6, 3.5)$, 
the sparticle spectra, the Higgs boson mass $m_h$, and the degree of 
tuning a $\mu$-term $|\Delta_\mu|\times 100(\%)$ 
are summarized in Tables~\ref{tab:LSP}, 
\ref{tab:sparticlemass}, and \ref{tab:higgs}. 
It is then satisfied by all the experimental lower bounds from 
the LHC experiments for the masses of all sparticles 
in Refs.~\cite{Aad:2014lra} and \cite{Beringer:1900zz} . 
In general, the SUSY flavor violations are dangerous in 
the gravity mediated SUSY-breaking scenario due to 
the flavor dependent interactions. 
In our setup, there are vanishing {\it A}-terms and no 
flavor dependent soft SUSY-breaking terms at the 
GUT scale because the moduli do not have the {\it F}-terms. 
Even if the moduli have the {\it F}-terms at the vacuum, 
the SUSY flavor violations can be suppressed from the 
structure of the $U(1)_{I'}$ charge assignments~\cite{Abe:2011rg}. 
Thus, there are no serious SUSY flavor violations; especially, 
the decay rates such as $\mu \rightarrow e\gamma$ and 
$b\rightarrow s\gamma$ evade the present 
limits~\cite{Adam:2013mnn,Amhis:2014hma}. 
\begin{table}[h]
\begin{center}
\begin{tabular}{|c|c||c|c|} \hline 
Sparticles&Mass[GeV] &Sparticles&Mass[GeV]\\ \hline
$\tilde{u}_1$ &2618&$\tilde{e}_1$ &3241\\ \hline
$\tilde{u}_2$ &2359&$\tilde{e}_2$ &2525\\ \hline
$\tilde{c}_1$ &2520&$\tilde{\mu}_1$ &2421\\ \hline
$\tilde{c}_2$ &2011&$\tilde{\mu}_2$ &2331\\ \hline
$\tilde{t}_1$ &1735&$\tilde{\tau}_1$ &2133\\ \hline
$\tilde{t}_2$ &974 &$\tilde{\tau}_2$ &1447\\ \hline
$\tilde{d}_1$ &2625 &$\tilde{\nu}_{e_1}$ &3240\\ \hline
$\tilde{d}_2$ &2620 &$\tilde{\nu}_{e_2}$ &415\\ \hline
$\tilde{s}_1$ &2522 &$\tilde{\nu}_{\mu_1}$ &2330\\ \hline
$\tilde{s}_2$ &2189 &$\tilde{\nu}_{\mu_2}$ &415\\ \hline
$\tilde{b}_1$ &2117 &$\tilde{\nu}_{\tau_1}$ &2132\\ \hline
$\tilde{b}_2$ &1724 &$\tilde{\nu}_{\tau_2}$ &415\\ \hline
$\tilde{\chi}_1^0$ &1723&$\tilde{\chi}_1^{\pm}$ &444\\ \hline
$\tilde{\chi}_2^0$ &1135&$\tilde{\chi}_2^{\pm}$ &1723\\ \hline
$\tilde{\chi}_3^0$ &448& & \\ \hline
$\tilde{\chi}_4^0$ &441& & \\ \hline
 \end{tabular}
 \caption{A typical sparticle spectra at the EW scale for the 
 reference point, $(r_1, r_2)=(6, 3.5)$. The subscripts denote 
the mass eigenvalues for the following: up ($\tilde{u}$), charm ($\tilde{c}$), 
top ($\tilde{t}$), down ($\tilde{d}$), strange ($\tilde{s}$), bottom ($\tilde{b}$) 
squarks, the scalar electron ($\tilde{e}$), muon ($\tilde{\mu}$), 
tauon ($\tilde{\tau}$), neutrino ($\tilde{\nu}$), the neutralino ($\tilde{\chi}$), 
and the chargino ($\tilde{\chi}^\pm$).}
  \label{tab:sparticlemass}
\end{center}
\end{table} 
\begin{table}
\begin{center}
\begin{tabular}{|c|c|c|c|} \hline 
$m_h$[GeV]&$m_H$[GeV]&$m_A$[GeV]&$m_{H\pm}$[GeV] \\ \hline
125.4&1423&1423&1425 \\ \hline
$\Delta_{\mu}^{-1}\times 100(\%)$&$M_1(m_Z)$[GeV] &$M_2(m_Z)$[GeV]&$M_3(m_Z)$[GeV] \\ \hline
2.1& 1133& 1719&1575 \\ \hline
 \end{tabular}
 \caption{The neutral and charged Higgs boson masses $m_h$, $m_H$, $m_A$, and $m_{H_{\pm}}$, the degree of tuning 
a $\mu$-term, $|\Delta_{\mu}|\times 100(\%)$, and 
the gaugino masses at the EW scale for the reference point 
$(r_1,r_2)=(6,3.5)$.}
  \label{tab:higgs}
\end{center}
\end{table}

\clearpage

\section{Conclusion}
\label{sec:conclusion}
In this paper, we proposed the gravitino dark matter in the gravity 
mediated SUSY-breaking scenario based on the $4$D ${\cal N}=1$ SUGRA. 
The nontrivial K\"ahler metric of the SUSY-breaking field induces 
the mass hierarchies between the gravitino and the other sparticles 
for any value of the {\it F}-term of the SUSY-breaking field. 
Especially, the small K\"ahler metric of the SUSY-breaking field 
leads to the stable gravitino of mass ${\cal O}(100)\,{\rm GeV}$ 
with TeV scale gauginos and sparticles which would be the typical 
features in the natural MSSM with gravity mediation, if the gauge 
kinetic functions and the kinetic terms of the matter fields satisfy certain conditions. (See Ref.~\cite{Kersten:2009qk} for the case of CMSSM.)
In the stable gravitino scenario, one can consider the low-scale 
SUSY without the cosmological gravitino problem, only if the NLSP 
decays do not spoil the success of BBN. 

As a concrete model, we considered the $5$D SUGRA model on $S^1/Z_2$. 
Since the successful inflation mechanism as well as the moduli 
stabilization have been realized in $5$D 
SUGRA~\cite{Maru:2003mq,Ade:2015lrj}, we 
have estimated the moduli and inflaton decays into the gravitino 
dark matter.  
Although the produced gravitinos via the moduli decays seem 
to be dangerous from the cosmological point of view, 
their decays can be suppressed only if the moduli do not have 
the {\it F}-terms. 
Such a situation can be applied in our model, because the moduli, 
inflaton, and stabilizer fields have supersymmetric masses at the vacuum. 
Even if the supersymmetry is broken in the SUSY-breaking sector, 
their {\it F}-terms are suppressed by the gravitino mass at the SUSY-breaking minimum. 
When the NLSP and NNLSP are taken as the sneutrino 
and Higgsino-like neutralino, the nonthermal productions of 
the gravitino are negligible due to the small thermal 
abundance of the Higgsino-like neutralino. 
The smallness of the thermal abundance of NNLSP 
also relaxes the constraints from the BBN~\cite{Kanzaki:2006hm,Feng:2004zu,Ishiwata:2007bt}, 
and at the same time, the amount of neutrinos via the sneutrino decay can be suppressed. 
Thus, the total relic abundance of the gravitino is approximated 
by the thermal abundance of it which depends on the gaugino 
masses. 
As pointed out in~\cite{Abe:2007kf}, the certain ratios of gaugino masses are 
also important to raise the Higgs boson mass in the MSSM without 
a severe fine-tuning. 
From Fig.~\ref{fig:higgsmugra}, it is found that certain 
ratios of gaugino masses are consistent with the relic abundance of 
the gravitino as well as the Higgs boson mass reported by the recent Planck 
and LHC data~\cite{Aad:2012tfa,Ade:2015lrj}.

In this paper, we focus on the $5$D SUGRA in order to show the 
realistic gravitino dark matter in the gravity mediation, and then 
the suppressed K\"ahler metric of the SUSY-breaking field is 
important to generate the mass hierarchies between the 
gravitino and other sparticles. 
When the $5$D SUGRA is derived as the effective theory of 
superstring theories on a warped throat and/or M-theory on the Calabi-Yau 
manifold~\cite{Lukas:1998yy}, the SUSY-breaking sector would be constructed 
from the gauge theory living on Dp-branes and/or NS5-branes. 
Especially, in the type II string, the visible and hidden 
sectors can be realized on the different D-branes which wrap 
the certain cycles in the internal manifold. 
In such cases, the different volumes of the internal cycles lead to the 
hierarchical K\"ahler metric between the SUSY-breaking 
field and matter fields in the visible sector. 

\subsection*{Acknowledgement}
The author would like to thank H.~Abe, T.~Higaki, J.~Kawamura, 
T.~Kobayashi, and Y.~Yamada for useful discussions and comments. 
H.~O. was supported in part by a Grant-in-Aid for JSPS Fellows 
No. 26-7296.

\appendix

\section{The F-terms of fields at the vacuum}
\label{app:oscvac}
In this appendix, we derive the {\it F}-terms of the moduli, stabilizer, 
and SUSY-breaking fields at the vacuum by employing the 
reference point method. As discussed in Sec.~\ref{subsec:moduli_stabilization}, 
when we expand the fields around the reference 
points given by Eqs.~(\ref{eq:refmoduli}) and (\ref{eq:modulivev}), 
$\phi\rightarrow \phi|_{\rm ref} 
+\delta \phi$, $\phi=T^{I'}, H_i, X$ with $I',i=1,2,3$, 
the K\"ahler metric with K\"ahler potentials~(\ref{eq:Kmo}) and (\ref{eq:KWX}) 
are expanded by
\begin{align}
K_{I\bar{J}} = K_{I\bar{J}}^{(0)} + K_{I\bar{J}}^{(1)},
\end{align}
where 
\begin{align}
K_{I\bar{J}}^{(0)}&=
\begin{pmatrix}
1/(2 {\rm Re}T^1)^2 & 0 & 0 & 0 & 0 & 0 & 0\\
0 & 1/(2 {\rm Re}T^2)^2 & 0 & 0 & 0 & 0 & 0\\
0 & 0 & 1/(2 {\rm Re}T^3)^2 & 0 & 0 & 0 & 0\\
0 & 0 & 0 & Z_{H_1} & 0 & 0 & 0\\
0 & 0 & 0 & 0 & Z_{H_2} & 0 & 0\\
0 & 0 & 0 & 0 & 0 & Z_{H_3} & 0\\
0 & 0 & 0 & 0 & 0 & 0 & Z_{X}-4|X|^2/\Lambda^2  \\
\end{pmatrix}
,
\end{align}
and
\begin{align}
K_{I\bar{J}}^{(1)}&=
\begin{pmatrix}
0 & 0 & 0 & a_{H_1} H_1 & 0 & 0 & a_X^1 X\\
0 & 0 & 0 & 0 & a_{H_2} H_2 & 0 & a_X^2 X\\
0 & 0 & 0 & 0 & 0 & a_{H_3} H_3 & a_X^3 X\\
a_{H_1} \bar{H}_1 & 0 & 0 & 0 & 0 & 0 & 0\\
0 & a_{H_2} \bar{H}_2 & 0 & 0 & 0 & 0 & 0\\
0 & 0 & a_{H_3} \bar{H}_3 & 0 & 0 & 0 & 0\\
a_X^1 \bar{X} & a_X^2 \bar{X} & a_X^3 X & 0 & 0 & 0 & 0 \\
\end{pmatrix}
,
\end{align}
in the field basis ($T^1,T^2,T^3,H_1,H_2,H_3,X$), with 
\begin{align}
&a_{H_i} \equiv \cfrac{\partial Z_{H_i}}{\partial T^{I'}} = \cfrac{1}{{\rm Re}T^{I'}}\left( e^{-2c_{H_i}{\rm Re}T^{I'}} - \cfrac{Z_{H_i}}{2}\right),\,\,\, (I'=i), \nonumber\\
&a_{X}^i \equiv \cfrac{\partial Z_{X}}{\partial T^{I'}} = \cfrac{c_X^i}{c_X \cdot {\rm Re}T}\left( e^{-2c_{X} \cdot ReT} - \cfrac{Z_{X}}{2}\right), \nonumber\\
\end{align}
and the inverse of the K\"{a}hler metric is given by
\begin{align}
&K^{T^{I'} \bar{T}^{I'}} \simeq (2{\rm Re}T^{I'})^2 +8{\rm Re}T^{I'}
{\rm Re}\,\delta T^{I'},\,\,\,
K^{T^{I'} \bar{H}_i} \simeq -A_{H_i} \delta H_i, \nonumber\\
&K^{T^{I'} \bar{X}} \simeq -A_{X}^i X -A_{X}^i \delta X,\,\,\,\,
K^{H_i \bar{H}_i} \simeq \cfrac{1}{Z_{H_i}} -\cfrac{2a_{H_i}}{(Z_{H_i})^2}
{\rm Re}\,\delta T^{I'},\nonumber\\
&K^{X \bar{X}} \simeq \cfrac{1}{Z_{X}} -\cfrac{2a_X^i}{(Z_X)^2}
{\rm Re}\,\delta T^{I'} +\cfrac{4}{\Lambda^2 (Z_X)^2} |\delta X|^2, 
\end{align}
where 
\begin{align}
&A_{H_i}\equiv (2 {\rm Re}T^{I'})^2 \cfrac{a_{H_i}}{Z_{H_i}},\,\,\,
A_{X}^i\equiv (2 {\rm Re}T^{I'})^2 \cfrac{a_{X}^i }{Z_{X}}.
\end{align}
Here and hereafter, we omit the subscript of $\phi$ at the reference 
point, that is, $\phi=\phi|_{\rm ref}$.  
From the relevant expansions in the scalar potential~(\ref{eq:scalarpo}) 
with the K\"ahler and superpotential~(\ref{eq:Kmo}), (\ref{eq:Wmo}), and (\ref{eq:KWX}), 
\begin{align}
D_{T^{I'}} W &\simeq K_{T^{I'}}w, \nonumber\\
              & + W_{T^{I'} H_i} \delta H^i +K_{T^{I'} {\bar T}^{I'}} 
w(\delta T^{I'}+\delta {\bar T}^{I'}) +K_{T^{I'}}W_X \delta X \nonumber\\
              &+W_{T^{I'} T^{I'} H_i}\delta T^{I'} \delta H_i +\sum_{J'=k} K_{T^{I'}} 
W_{T^{J'} H_k} \delta T^k \delta H^k, \nonumber\\              
D_{H_i} W &\simeq W_{T^{I'} H_i} \delta T^{I'} +K_{H_i {\bar H}_i} w\delta {\bar H}_i
+ \cfrac{W_{T^{I'} T^{I'} H_i}}{2}  (\delta T^{I'})^2, \nonumber\\
D_{X} W &\simeq W_X + K_{X {\bar X}} w\delta {\bar X} +\frac{1}{2}\partial_X 
(K_{X\bar{X}})w(2|\delta X|^2 +(\delta X)^2+(\delta \bar{X})^2), \nonumber\\ 
W &\simeq w + W_X \delta X +\sum_{I'=i}^3 W_{T^{I'} H_i} \delta T^{I'} \delta H_i, \nonumber\\
K &\simeq \sum_{I'=1}^3\left( -{\rm ln} ({\rm Re}\,T^{I'}) 
-\cfrac{{\rm Re}\,\delta T^{I'}}{{\rm Re}\,T^{I'}} + 
\cfrac{1}{2}\left(\cfrac{{\rm Re}\,\delta T^k}{{\rm Re}\,T^{I'}}\right)^2 \right),
\end{align}
we obtain the scalar potential at the second order $\delta \phi$, 
\begin{align}
V &\simeq \frac{W_X^2}{Z_X} -2wW_X (\delta X +\delta\bar{X}) 
-\sum_{I'=i}(2\text{Re}T^{I'})wW_{T^{I'}H_i} (\delta H_i +\delta\bar{H}_i) \nonumber\\
&\hspace{13pt}+\frac{4w^2}{\Lambda^2 Z_X^2} |\delta X|^2 +\sum_{I'=i} \frac{W_{T^{I'}H_i}^2}{Z_{H_i}}|\delta T^{I'}|^2 +\sum_{I'=i}^3 (2\text{Re}\,T^{I'})^2
W_{T^{I'}H_i}^2 |\delta H_i|^2 \nonumber\\
&\hspace{13pt}+\sum_{I'=i}^3(-2\text{Re}\,T^{I'} wW_{T^{I'}T^{I'}H_i}+wW_{T^{I'}H_i}) (\delta T^{I'}\delta H_i +\delta\bar{T}^{I'}\delta\bar{H}_i) \nonumber\\
&\hspace{13pt}+\sum_{I'=i}\frac{A_{H_i}}{2\text{Re}T^{I'}} wW_{T^{I'}H_i}
(\delta T^{I'}\delta\bar{H}_i +\delta\bar{T}^{I'}\delta H_i) 
-\sum_{i=1}^3 wW_{T^{I'}H_i}(\delta T^{I'} +\delta\bar{T}^{I'})(\delta H_i 
+\delta\bar{H}_i) \nonumber\\
&\hspace{13pt}+\sum_{I'=i}^3 2\text{Re}\,T^{I'} W_X W_{T^{I'}H_i}(\delta H_i
\delta\bar{X} +\delta\bar{H}_i\delta X) \nonumber\\
&\hspace{13pt}+\sum_{I'=i}^3\sum_{J'=1}^3 \frac{T^{I'}+\bar{T}^{I'}}{T^{J'}+\bar{T}^{J'}}wW_{T^{I'}H_i}(\delta T^{J'} +\delta\bar{T}^{J'})(\delta H_i +\delta\bar{H}_i).
\label{scalarpotential}
\end{align}
Finally, the extremal conditions for the relevant fields lead to the 
following variations of them:
\begin{align}
&\delta H_i \simeq \frac{w}{2\text{Re}T^{I'} W_{T^{I'}H_i}}\sim 
{\cal O}\left(\frac{m_{3/2}}{m_{H_i}}\right),\,\,\,
\delta X\simeq \left(\frac{\Lambda^2 Z_X^2}{4w^2}\right) 5wW_X,
\nonumber\\
&\delta T^{I'}\simeq \left(\frac{w}{W_{T^{I'}H_i}}\right)^2 Z_{H_i}\left( \frac{1+A_{H_i}K_{T^{I'}}}{2\text{Re}T^{I'}} +\frac{W_{T^{I'}T^{I'}H_i}}{W_{T^{I'}H_i}} 
-\frac{3}{\text{Re}T^{I'}}\right) \sim 
{\cal O}\left(\frac{m_{3/2}}{m_{T^{I'}}}\right),
\label{variation}
\end{align}
and their {\it F}-terms become
\begin{align}
&\sqrt{K_{T^{I'}\bar{T}^{I'}}} F^{T^{I'}} = -e^{K/2} \sqrt{K_{T^{I'}\bar{T}^{I'}}} 
K^{T^{I'}\bar{J}} \overline{D_JW} \sim O\left(\frac{m_{3/2}^3}{m_{T^{I'}}^2}\right),\nonumber\\
&\sqrt{K_{H_i\bar{H}_i}} F^{H_i} = -e^{K/2} \sqrt{K_{H_i\bar{H}_i}} K^{H_i\bar{J}} 
\overline{D_JW} \sim O\left(\frac{m_{3/2}^3}{m_{H_i}^2}\right),\nonumber\\
&\sqrt{K_{X\bar{X}}} F^X \simeq -e^{K/2} \sqrt{K_{X\bar{X}}} K^{X\bar{X}} D_XW \simeq \frac{-W_X}{(\text{Re}T^1\text{Re}T^2\text{Re}T^3)^{1/2}Z_X^{1/2}},
\end{align}
where 
\begin{align}
&D_{T^{I'}} W = \text{min}\left(O\left(\frac{m_{3/2}^3}{m_{T^{I'}}^2}\right), O\left(\frac{m_{3/2}^3}{m_{X}^2}\right) \right),\,\,\, (I'=1,2),
\nonumber\\
&D_{T^3} W = O\left(\frac{m_{3/2}^3}{m_{T^3}^2}\right), \;\;D_{H_i} W = O\left(\frac{m_{3/2}^2}{m_{H_i}}\right), (i=1,2,3),\,\,\,
D_X W\simeq  \nu.
\end{align}
We also numerically checked these results, and then their {\it F}-terms 
can be suppressed by the tiny mass of the gravitino.

\section{The minima of fields during the inflation}
\label{app:oscinf}
By contrast, the minima of fields during the inflation are different from 
those at the true vacuum. 
In this section, we derive the minima of fields by employing 
the reference point method. 
The reference points of fields during the inflation are chosen in 
the same way as those at the vacuum.
Similarly, we expand the fields except for the inflaton Re\,$T^3$ 
around the reference points given by Eqs.~(\ref{eq:refmoduli}) and (\ref{eq:modulivev}), $\phi\rightarrow \phi|_{\rm ref} 
+\delta \phi$, $\phi=T^{I'},{\rm Im}\,T^3, H_i, X$ with $I'=1,2$, $i=1,2,3$. 
It is then supposed that $H_3$ is fixed at the origin due to the 
Hubble-induced mass. 
From the scalar potential~(\ref{eq:scalarpo}) with the K\"ahler and 
superpotentials~(\ref{eq:Kmo}), (\ref{eq:Wmo}), and (\ref{eq:KWX}) 
given by the following expansions:
\begin{align}
D_{T^{I'}} W &\simeq K_{T^{I'}} w+ W_{T^{I'} H_i} \delta H_i +K_{T^{I'} {\bar T}^{I'}} 
w(\delta T^{I'} +\delta {\bar T}^{I'}) +K_{T^{I'}}(W_{H_3} \delta H_3 +W_X \delta X) \nonumber\\
              & + K_{T^{I'} {\bar T}^{I'}} W_{H_3} (\delta T^{I'}+\delta {\bar T}^{I'}) \delta H^3 +W_{T^{I'} T^{I'} H_i}\delta T^{I'} \delta H_i +K_{T^{I'}}\sum_{J'=j}^3 W_{T^{J'} H^j} \delta T^{J'} \delta H^j, \nonumber\\                   
D_{H_1} W &\simeq W_{T^1 H_1} \delta T^1 +K_{H_1 {\bar H}_1} w\delta {\bar H}_1+ K_{H_1 {\bar H}_1} W_{H_3} \delta {\bar H}_1 \delta H_3, \nonumber\\
D_{H_2} W &\simeq W_{T^2 H_2} \delta T^2 +K_{H_2 {\bar H}_2} w\delta {\bar H}_2+ K_{H_2 {\bar H}_2} W_{H_3} \delta {\bar H}_2 \delta H_3, \nonumber\\ 
D_{H_3} W &\simeq W_{H_3}+ W_{T^3 H_3} \delta T^3 +K_{H_3 {\bar H}_3} W\delta {\bar H}_3+\partial_{T^3}(K_{H_3 {\bar H}_3})w (\delta T^3+\delta {\bar T}^3)\delta {\bar H}_3, \nonumber\\
              &+K_{H_3 {\bar H}_3}  (W_{H_3} |\delta H_3|^2 +W_X \delta {\bar H}_3 \delta X)+\cfrac{W_{T^3 T^3 H_3}}{2}  (\delta T^3)^2,\nonumber\\ 
D_{X} W &\simeq W_{X} +K_{X {\bar X}} w\delta {\bar X}+\frac{1}{2}\partial_X (K_{X\bar{X}})w(2|\delta X|^2 +(\delta X)^2+(\delta \bar{X})^2)+ K_{X {\bar X}} W_{H_3} \delta H_3 \delta {\bar X}, \nonumber\\ 
W &\simeq w + W_X \delta X +W_{H_3}\delta H_3+\sum_{I'=i}^3 W_{T^{I'} H_i} \delta T^{I'} \delta H_i, \nonumber\\
K &\simeq \sum_{I'=i}^3\left( -{\rm ln} {\rm Re}\,T^{I'} -\cfrac{{\rm Re}\,\delta T^{I'}}{{\rm Re}\,T^{I'}} + \cfrac{1}{2}\left(\cfrac{{\rm Re}\,\delta T^{I'}}{{\rm Re}\,T^{I'}}\right)^2\right) +\sum_i Z_{H_i} |\delta H_i|^2  +Z_X |\delta X|^2,
\end{align}
we obtain the extremal conditions for the relevant fields, and then their variations become
\begin{align}
&\delta \tau^1=\delta \tau^2=\delta \tau^3=\delta k_1=\delta k_2=\delta k_3= \delta y=0, \nonumber \\ 
&\delta \sigma^{I'} \sim O\left( \frac{Z_{H_i}}{W_{T^{I'}H_i}^2 \text{Re}\,T^{I'}} 
\frac{|W_{H_3}|^2}{Z_{H_3}}\right) \simeq O\left(\left( \frac{H_{\rm inf}}{m_{T^{I'}}}\right)^2\right),\,\,\, (I'=1,2),\nonumber \\
&\delta h_i \sim O\left( \frac{K_{T^{I'}}w }{W_{T^{I'}H_i} (2\text{Re}T^{I'})^2 } \right) 
\simeq O\left( \frac{m_{3/2}}{m_{H_i}}\right), \,\,\,(i=1,2),
\nonumber\\
&\delta h_3 \sim O\left( \frac{w}{W_{H_3}}\right) 
\simeq O\left(\frac{m_{3/2}}{m_{H_3}}\right),
\nonumber \\
&\delta x \sim O\left( \frac{\Lambda^2 Z_X^2}{4W_X^2} W_X w\right) \simeq O\left(\left( \frac{m_{3/2}}{m_{X}}\right)^2\right), 
\end{align}
where 
\begin{align}
& \delta T^{I'} \equiv\delta \sigma^{I'} + i\delta \tau^{I'}, \;\; 
\delta H_i \equiv \delta h_i +i\delta k_i, \;\; \delta X \equiv \delta x+i\delta y,
\end{align}
with $I',i=1,2,3$.

\end{document}